\pdfoutput=1
\documentclass[11pt]{article}
\usepackage{jheppub}

\usepackage{customprelude}

\def\be{\begin{equation}}
\def\ee{\end{equation}}
\def\ba{\begin{array}}
\def\ea{\end{array}}
\def\ad3{\ensuremath{\overline{\rm{D3}}}}
\def\vo{\mathcal{V}}

\title{\centering Global String Embeddings for the Nilpotent Goldstino}

\author[a]{I\~naki Garc\'ia-Etxebarria,}
\author[b,c]{Fernando Quevedo,}
\author[d,e]{and Roberto Valandro}

\affiliation[a]{Max Planck Institute for Physics, F\"ohringer Ring 6, 80805 Munich, Germany}
\affiliation[b]{DAMTP, CMS, University of Cambridge, Wilberforce Road,
  Cambridge, CB3 0WA, UK}
\affiliation[c]{ICTP, Strada Costiera 11, 34151 Trieste, Italy}
\affiliation[d]{Dipartimento di Fisica, Universit\`a di Trieste, Strada Costiera 11, 34151 Trieste, Italy}
\affiliation[e]{INFN, Sezione di Trieste, Via Valerio 2, 34127 Trieste, Italy}

\emailAdd{inaki@mpp.mpg.de}
\emailAdd{f.quevedo@damtp.cam.ac.uk}
\emailAdd{roberto.valandro@ts.infn.it}

\abstract{We discuss techniques for embedding a nilpotent Goldstino
  sector both in weakly coupled type IIB compactifications and
  F-theory models at arbitrary coupling, providing examples of both
  scenarios in semi-realistic compactifications. We start by showing
  how to construct a local embedding for the nilpotent Goldstino in
  terms of an anti D3-brane in a local conifold throat, and then
  discuss how to engineer the required local structure in globally
  consistent compact models. We present two explicit examples, the
  last one supporting also chiral matter and K\"ahler moduli
  stabilisation.}

\begin{document}


\makeatletter
\let\old@fpheader\@fpheader
\renewcommand{\@fpheader}{\old@fpheader
\hfill MPP-2015-311

\hfill DAMTP-2015-91}
\makeatother

\maketitle


\section{Introduction}

$\mathcal{N}=1$ supergravity theories coupled to matter have been studied for more than 30 years. The combination of supersymmetry and chirality makes them one of the most interesting effective field theories (EFT) that can address unsolved issues of particle physics. They are also the natural effective field theories that represent the dynamics of chiral low-energy string modes upon compactifications on Calabi-Yau (CY) spaces (where in fact supersymmetry  plays an important role for having proper control on the  EFT). Matter is usually represented by chiral superfields and supersymmetry is linearly realised. But further constraints may be imposed on the chiral superfields that can furnish non-linear representations of supersymmetry. 

The simplest case is a superfield $X$ satisfying a nilpotent condition $X^2=0$. Such a superfield $X$ has only one propagating component, 
that can be identified with the goldstino arising from spontaneously supersymmetry breaking at higher scales. Since the scalar component of $X$ is a bilinear of the fermion component that gets zero vev and the most  general superpotential is linear in $X$, its contribution to the total scalar potential is a positive definite term $\delta V\propto |\partial W/\partial X|^2$ that can be used to lift the minimum of the scalar potential and potentially lead to de Sitter vacua \cite{Bergshoeff:2015tra,Dudas:2015eha,Antoniadis:2015ala,Hasegawa:2015bza,Kallosh:2015tea,Dall'Agata:2015zla,Schillo:2015ssx,Kallosh:2015pho}.

In string compactifications it has recently been realised that a
nilpotent superfield might capture the low-energy physics representing
the presence of an anti-D3-brane at the tip of a throat
\cite{Kallosh:2014wsa,Bergshoeff:2015jxa,Kallosh:2015nia,Aparicio:2015psl}
(see \cite{Bandos:2015xnf} for a complementary approach).  This setup
was the basic ingredient in the original proposal of KKLT
\cite{Kachru:2003aw} to obtain de Sitter space in flux
compactifications with stabilised moduli \cite{Giddings:2001yu}. In
\cite{Kallosh:2015nia} explicit string realisations were found in
which the presence of an anti-D3-brane leaves the goldstino as the
only low-energy degree of freedom, justifying the use of a nilpotent
superfield $X$ to describe the EFT.  In particular this is true if the
anti-D3-brane is on top of an O3-plane at the tip of a warped throat
with (2,1) three-form fluxes.  The constructions presented in
\cite{Kallosh:2015nia} were at the local level, and constructing a
fully-fledged compact string construction with a nilpotent goldstino
was left as an open challenge.

In this article we address the open issue of embedding the local setup
of \cite{Kallosh:2015nia} in a compact Calabi-Yau. We first analyse in
a systematic way the local approaches to obtain a goldstino in local
conifold-like geometries obtained by orientifolded conifolds, refining
and generalising the analysis in \cite{Kallosh:2015nia}. Very
importantly for our purposes of finding global embeddings, and
contrary to what was claimed in \cite{Kallosh:2015nia}, we find that
already the standard conifold singularity
\cite{Candelas:1989js,Klebanov:1998hh} can support an orientifold
involution necessary to produce an O3-plane at the tip of the
throat. This O3-plane is necessary to obtain the spectrum encoded in
the nilpotent superfield. We show that, deforming the conifold
singularity leads to two O3-planes sit on the blown-up three-sphere at
the tip of the throat. By a field theory analysis, based on probe
D3-branes, we identified the O-plane type, finding that for our choice
of involution the two O3-planes are either both $O3^{-}$ or both
$O3^{+}$. We also verify our conclusions by comparing the results with
the T-dual type IIA setup.

\begin{figure}[!t]
\centering
\includegraphics[width=10.0cm]{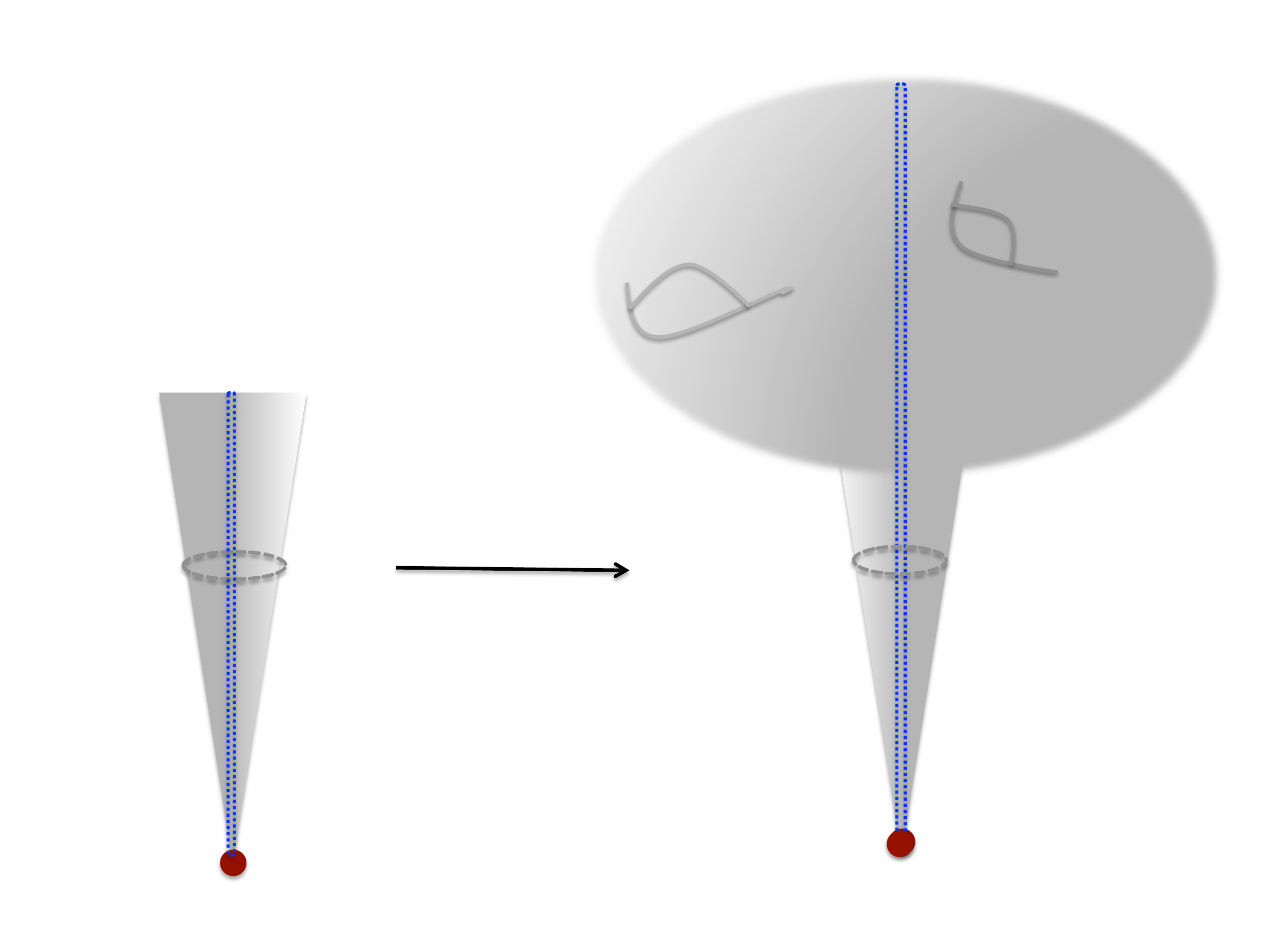}
\caption{From local to global orientifold realisation of the anti-D3-brane at the tip of orientifolded conifold threaded by three-form fluxes on two dual three-cycles}
\label{Fig1}
\end{figure}

After  settling the local setup, we
proceed to embed it in
globally consistent compact string theory backgrounds, as shown
schematically in figure~\ref{Fig1}.
We followed two strategies to do this. First we construct a compact
non-CY threefold with the wanted properties, i.e. it has a local patch that behaves as 
the deformed conifold geometry and an involution that restricts on the local patch as the
involution studied previously. Then, in the F-theory context we use this manifold to  create an elliptically fibred 
Calabi-Yau fourfold. The weak coupling Sen limit allows then to construct a Calabi-Yau three-fold
with the wanted features.

The second strategy is based on searching for suitable manifolds among the Calabi-Yau 
hypersurfaces in toric varieties \cite{Kreuzer:2000xy}. We look for spaces and involutions that produce
more than one O3-plane. Among these we choose the one where there 
is a complex structure deformation that leads two O3-planes on top
of the same point, and at the same time produces a conifold singularity at this point.
Then deforming back to a smooth CY, we obtain the wanted configuration.
By these methods we find two explicit examples of Calabi-Yau where the nilpotent goldstino can be embedded.

Independent of the goldstino representation, it is important to
emphasise that despite the fact that the KKLT proposal for de Sitter
uplift was presented more than 10 years ago, the explicit realisation
of the anti-D3-brane uplift in a globally defined compactification,
including potentially chiral matter had, to the best of our knowledge, not been
achieved so far. It is  one of the motivations for the current
article to fill this gap.

\medskip

This article is organised as follows. In section~\ref{Sec:AntiD3} we
recall the basic issues regarding the $\ad3$ brane uplift and its
representation in an EFT by nilpotent
superfields. Section~\ref{sec:conifold-intro} is devoted to addressing
in a systematic way the local realisation of an $\ad3$ sitting on top
of orientifold plane configuration O3 at the tip of a deformed and
orientifolded Klebanov-Strassler (KS) throat.  Finally in
section~\ref{sec:global-embeddings} we address the main goal of the
article which is to embed the local constructions into compact CY
backgrounds. We present two concrete examples. In the first example we
illustrate how to construct models with the right local structure
basically from scratch. It turns out that F-theory provides an
efficient way of building such models. The second example is in fact a
Calabi-Yau that had already been studied in the model building context
before. We show that it has the right local structure in order to
admit a nilpotent Goldstino sector. We end with the conclusions in
section~\ref{Sec:conclus}.

\section{Anti-D3-branes and nilpotent goldstino}\label{Sec:AntiD3}

In type IIB string theory has RR and NSNS three forms field strength, encoded into the complex three-form $G_3$, can thread quantised fluxes on the non-trivial 3-cycles of Calabi-Yau compactifications. Their impact is to fix the corresponding complex structure moduli and at the same time inducing a warp factor $e^{2D}$ in the background metric:
\begin{equation}
ds^2=e^{2D} ds_4^2+ e^{-2D} ds_{CY}^2 \:.
\end{equation}
One can write the (internal coordinate dependent) warp factor such as 
$e^{-4D}=1+\frac{e^{-4A}}{\vo^{2/3}}$. A large warped region, called warped throat, 
is made up of points where $e^{-2D}\gg \vo^{1/3}$.
Typically these throats arise around deformed conifold singularities.
At the tip of the throat one finds the blown-up three-sphere. The warp factor at the tip
depends on the flux numbers $K,M$ (that are the integrals of the three-form fluxes
on the three-sphere and its dual three-cycle)~\cite{Giddings:2001yu}:  
$e^{4A_0}= e^{-8\pi K/3g_sM}\ll 1$.
Depending on the relative value of the integer fluxes ($K,M$) the corresponding 
warp factor may give rise to a long throat. 

These fluxes combined with non-perturbative effects are enough to fix
all geometric moduli and the dilaton but usually lead to a negative
vacuum energy and therefore anti de Sitter space.  Adding an anti-D3-brane 
at the tip of a throat adds a positive component to the vacuum
energy and can uplift the minimum to de Sitter space. Notice that the
anti-D3-brane will naturally minimise the energy by sitting precisely
at the tip of a throat in which the warp factor provides the standard
redshift factor to reduce the corresponding scale. Furthermore, this
redshift is crucial for the effective field theory describing the
presence of the anti-D3-brane to be well defined since the
contribution to the energy of the anti-D3-brane is
\cite{Kachru:2003sx}
\begin{equation}
\delta V = M_{ws}^4=\vo^{2/3}e^{4A_0}M_s^4\sim \frac{e^{4A_0}M_p^4}{\mathcal{V}^{4/3}}\ll M_s^4
\end{equation}
where $M_{ws}$ is the warped string scale, $e^{4A_0}$ the warp factor at the tip of the throat and $\mathcal{V}$ the volume of the extra dimensions. $M_s$ and $M_p$ are the string and Planck scale respectively. Since the effective field theory is only valid at scales
smaller than the string scale $M_s^4$ a hierarchically small warp factor is needed to have a consistent field theory description of the anti-D3-brane.

On an independent direction constrained superfields have been considered on and off over the years \cite{Rocek:1978nb,Ivanov:1978mx,Lindstrom:1979kq,Casalbuoni:1988xh,Komargodski:2009rz}. A chiral nilpotent superfield $X$ can be written as
\be
X(y,\theta)= X_0(y)+\sqrt{2}\psi(y) \theta + F(y) \theta \overline{\theta},
\ee
with, as usual,  $y^\mu=x^\mu+i\theta\sigma^\mu\overline{\theta}$. The nilpotent condition $X^2=0$ implies $2X_0=\psi\psi/F$ and therefore does not propagate. It furnishes a non-linear representation of supersymmetry with a single propagating component, the goldstino $\psi$.

 For a string compactification after fixing the dilaton and complex structure moduli the K\"ahler potential for K\"ahler moduli and nilpotent goldstino can be written as
\be
K=-2\log \mathcal{V} + \frac{\alpha}{\mathcal{V}^n} XX^* \:,
\ee
while the superpotential is
\be
W=\rho X+ W_0 \:,
\ee
where we have used the fact that higher powers of $X$ are zero because of the nilpotency of $X$.
The scalar potential contribution of $X$ is then
\be
\delta V_X=M_p^4e^KK^{-1}_{XX^*}\left|\frac{\partial W}{\partial X}\right|^2=\frac{M_p^4|\rho|^2}{\alpha\vo^{2-n}} \:,
\ee
which agrees with the KKLMMT \cite{Kachru:2003sx} result above for $n=2/3$ with the warp factor being reproduced by
$|\rho|^2/\alpha$ \cite{Kallosh:2015nia,Aparicio:2015psl}. 

Another effect of the three form fluxes $G_3$ is to give mass to some of the $\ad3$ brane states. One $\ad3$  brane by itself carries the degrees of freedom of an $\mathcal{N}=4$ vector multiplet. In the presence of supersymmetry preserving $(2,1)$ ISD fluxes the scalar fields inside the anti-D3-brane get massive, consistent with the fact that the $\overline{\rm{D3}}$ gets fixed at the tip of the throat. Fluxes also give mass to three of the four $\mathcal{N}=4$ fermions by the couplings $G_3\lambda\lambda$. This is through the coupling $\bf{10}\cdot \overline{\bf 4}\cdot \overline{\bf 4}$ in terms of representations of $SO(6)$ once they are decomposed in terms of $SU(3)\times U(1)$ representations relevant for $\mathcal{N}=1$ supersymmetry. Therefore $(2,1)$ fluxes leave only a $U(1)$ gauge field and one single fermion (goldstino) in the massless spectrum. 

In order to have only the goldstino in the spectrum and justify the use of the nilpotent $X$ superfield we need to project out the gauge field by orientifolding. Orientifold involutions are a basic component of type IIB compactifications. Having the action of the orientifold involution such that the tip of the throat coincides with the fixed point of the orientifold needs a detailed analysis that was started in reference \cite{Kallosh:2015nia}. We reconsider the local constructions in the next section, extending the analysis of  \cite{Kallosh:2015nia}, before embedding them in global constructions.

\section{The conifold embedding of the nilpotent Goldstino}
\label{sec:conifold-intro}

The local model of interest will be an isolated orientifold of the
conifold, which we parametrise by the equation
\begin{equation}
  \label{eq:singular-conifold}
  xy = zw
\end{equation}
in $\bC^4$, with a singularity at $x=y=z=w=0$. The deformed version of
the conifold is given by
\begin{equation}
  \label{eq:deformed-conifold}
  zw = xy + t^2\, .
\end{equation}
For simplicity we will often take $t\in \bR$.

\begin{figure}
  \centering
  \begin{subfigure}[t]{0.3\textwidth}
    \centering
    \includegraphics[height=5cm]{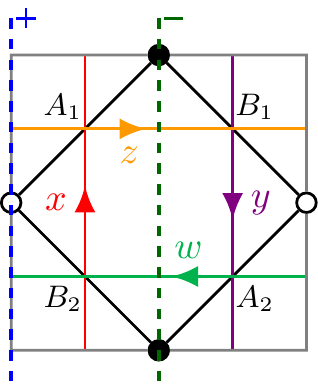}
    \caption{Dimer model.}
    \label{sfig:orientifolded-conifold-dimer}
  \end{subfigure}
  \hspace{1cm}
  \begin{subfigure}[t]{0.6\textwidth}
    \centering
    \includegraphics[width=0.8\textwidth]{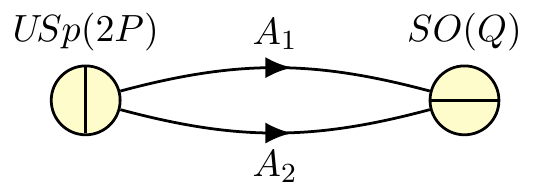}
    \caption{Quiver.}
    \label{sfig:orientifolded-conifold-quiver}
  \end{subfigure}
  \caption{\subref{sfig:orientifolded-conifold-dimer} Dimer model for
    the orientifold of the conifold that we are considering. The
    dashed lines indicate the dimer involution that we are considering
    (a line orientifold, in the nomenclature of
    \cite{Franco:2007ii}). The solid lines denote the four elementary
    mesons. We have also named the bifundamentals as in the
    text. \subref{sfig:orientifolded-conifold-quiver} The
    corresponding quiver. We have denoted the bifundamentals with
    arrows to indicate that they are $\cN=1$ chiral multiplets, but
    they live in real representations so the orientation of the arrow
    is immaterial.}
  \label{fig:orientifolded-conifold}
\end{figure}

We are interested in an orientifold action with geometric part acting
as
\begin{equation}
  \label{eq:xyzw-involution}
  \sigma\colon (x,y,z,w)\to (y,x,-z,-w)\, .
\end{equation}
In the $z_4\neq 0$ patch (and similarly for other patches) the
holomorphic three form for the conifold can be written as
\begin{equation}
  \Omega = \frac{dz_{1}\wedge dz_2 \wedge dz_3}{z_4}
\end{equation}
which transforms under~\eqref{eq:xyzw-involution} as
$\Omega\to -\Omega$, as befits an orientifold compatible with the
presence of D3-branes. Acting on the singular
conifold~\eqref{eq:singular-conifold} the
involution~\eqref{eq:xyzw-involution} leaves the origin $x=y=z=w=0$
fixed, while acting on the deformed
conifold~\eqref{eq:deformed-conifold} it leaves two fixed points at
$(x,y,z,w)=(\lambda,\lambda,0,0)$ with $\lambda=\pm it$ fixed. The
brane tiling and corresponding quiver for the theory of fractional
branes on the orientifolded singularity can be determined using the
techniques in \cite{Franco:2007ii}, or more directly via our explicit
type IIA construction below.

As is well known, in the absence of the orientifold the deformation of
the conifold takes place dynamically due to confinement in the brane
system \cite{Klebanov:2000hb}. The same is true in the presence of the
orientifold. Our goal in this section will be to clarify various
aspects of the dynamics of this orientifolded configuration. Most
importantly for our purposes, we will determine which type of
orientifold fixed plane arises after confinement, which we need to
know in order to construct explicit embeddings of the nilpotent
goldstino.\footnote{The problem of determining the orientifold charges
  was already studied in \cite{Ahn:2001hy,Imai:2001cq}. It was claimed
  in those papers that the orientifold planes appearing in the
  deformed description have opposite NSNS charge. We find instead
  (from various viewpoints) that the orientifold planes arising from
  confinement have the same NSNS charge.}

We will describe the physics of branes in type IIB language momentarily,
but we first discuss the physics of the type IIA dual, since it is clearer
in many respects.

\subsection{Type IIA perspective}
\label{sec:IIA-interpretation}

Let us start by reviewing well known facts about T-duality on the
conifold.\footnote{A more detailed discussion of the duality map can
  be found in \cite{Uranga:1998vf,Dasgupta:1998su}.} The singular
conifold $xy=zw$ has a $U(1)$ symmetry
\begin{equation}
  \label{eq:conifold-U(1)}
  (x,y,z,w) \to (e^{i\alpha}x, e^{-i\alpha}y, z, w)\, .
\end{equation}
The full symmetry group is $SU(2)\times SU(2)\times U(1)$, as is well
known \cite{Klebanov:1998hh}, but we focus on this subgroup for
convenience. We can view~\eqref{eq:conifold-U(1)} as a $\bC^*$
fibration over $(z,w)$, with the $\bC^*$ fibre constructed as the
hypersurface $xy=zw$ in the $(x,y)$ ambient $\bC^2$. The $\bC^*$ fibre
becomes singular at $\{z=0\}\cap\{w=0\}$. Fixing a finite radius at
infinity, we can T-dualise along this isometry and obtain a IIA dual
on $\bR^9\times S^1$, in the presence of NS5 branes located where the
$\bC^*$ fibre (or equivalently the $U(1)$
action~\eqref{eq:conifold-U(1)}) degenerates,
i.e. $\{z=0\}\cup\{w=0\}$. The position of the NS5 branes on the fibre
directions depends on the value of the $B$-field across the $\bP^1$
cycle in the resolved description of the conifold. For concreteness,
we label the coordinates as $x^i$, with $x^0,\ldots,x^3$ the four
Minkowski directions, $z=x^4+ix^5$, $w=x^8+ix^9$ and $x^6\in S^1$ the
direction on which we T-dualise. We have
\begin{equation}
  \begin{array}{c|cccccccccc}
    & 0 & 1 & 2 & 3 & 4 & 5 & 6 & 7 & 8 & 9\\
    \hline
    \text{NS5} & - & - & - & - & - & - \\
    \text{NS5}' & - & - & - & - & & & & & - & -\\
    \text{D4} & - & - & - & - & & & -
  \end{array}
\end{equation}
We have also indicated the D4-branes appearing from dualising a D3-brane at
the conifold. Fractional D3-branes are D4-branes ending on the NS5 and
NS5$'$ branes, instead of wrapping fully around the $x^6$ direction.

For the purposes of relating the IIA and
IIB pictures we write local coordinates $r,s$ for the $\bC^*$
\begin{equation}
  s+ir = \frac{1}{4\pi i}\log\left(\frac{x}{y}\right)\, .
\end{equation}
The $U(1)$ isometry acts by shifts on the periodic coordinate $s$,
leaving $r$ invariant. (We have introduced an extra factor of
$\frac{1}{2}$ so that $s\to s+1$ as we act with a full $U(1)$
rotation.) For finite asymptotic radius of the $\bC^*$ we have, far
enough from the core, a flat $\bR\times S^1$ geometry parametrised by
$(r,s)$. T-duality in this asymptotic region then acts on $s$ only, so
we identify $r\cong x^7$, and $s$ and $x^6$ are coordinates on the
T-dual circles.

The complex deformation of the conifold has equation $zw=xy+t^2$. For
simplicity we will take $t\in\bR$. Clearly the
isometry~\eqref{eq:conifold-U(1)} is still there, so we can still
T-dualise. The picture is similar, but now the two NS5 branes
recombine into the smooth 2-cycle $zw=t^2$.

\medskip

The previous discussion has nothing which is not well known. Let us
now orientifold the configuration, and see what we obtain. The
orientifold action of interest to us is given by
$(x,y,z,w)\to (y,x,-z,-w)$ in \eqref{eq:xyzw-involution}.  Exchanging $x$ with $y$, the action on
the local coordinates is $(r,s)\to (-r,-s)$.  Upon T-duality this maps to $x^7\to -x^7$. 
Together with the sign change in
$(z,w)$, this gives precisely an O4-plane wrapping $x^6$, as
expected. Recalling that the orientifold type changes when crossing a
NS5 brane, we find a $\Sp\times SO$ structure for the gauge algebra on
the branes, as in figure~\ref{fig:orientifolded-conifold}.

\begin{figure}
  \centering
  \includegraphics[width=\textwidth]{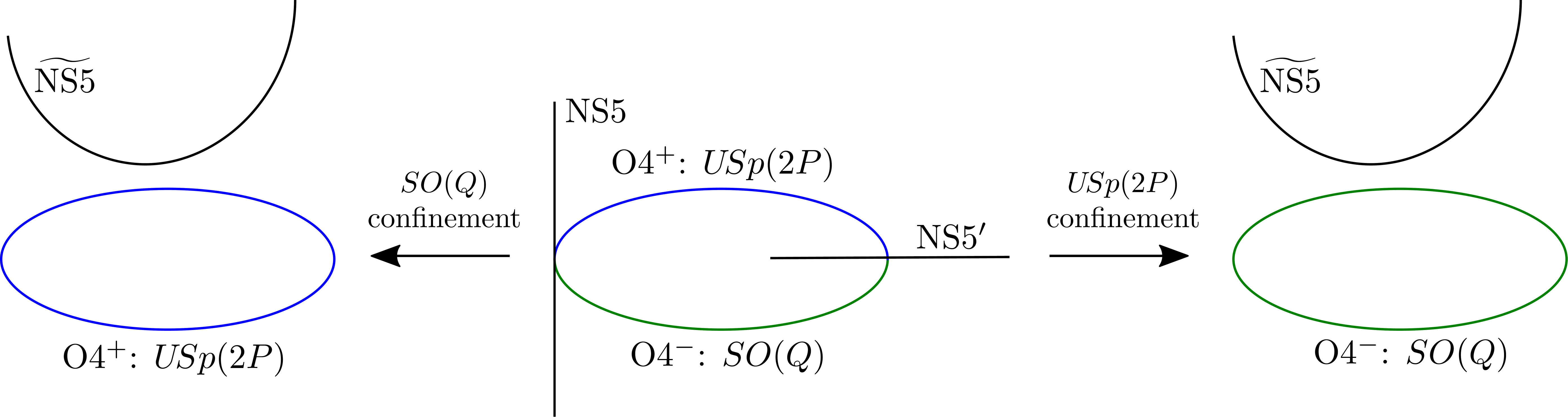}
  \caption{Confinement on the orientifolded conifold from the IIA
    perspective. The central Hanany-Witten \cite{Hanany:1996ie}
    configuration is the classical description. Confinement on the
    $SO(Q)$ factor (left diagram) corresponds to joining together the
    NS5 and NS5$'$ branes by shrinking to zero size the $SO(Q)$ side
    (i.e. the side with the O4$^-$ plane), and then recombining the
    two intersecting NS5 branes into the recombined object
    $\widetilde{\text{NS5}}$, which does not intersect the remaining
    O4$^+$ plane. Similarly confinement of the $\Sp(2P)$ factor
    (right) leads to a O4$^-$ plane after confinement. In either case,
    we observe that the remaining O4-plane has sign opposite to
    the O4-plane on the gauge factor giving rise to confinement.}
  \label{fig:IIA-confinement}
\end{figure}

Now we do the geometric deformation. There are two key facts to
observe: the locus $\{zw=t^2\}$ wrapped by the NS5 maps to itself
under $(z,w)\to(-z,-w)$, but it does so without any fixed points. So
the recombined NS5 does not intersect the O4. The two fixed points of
the deformed conifold in the $(x,y,z,w)$ coordinates are at
$(x,x,0,0)$, with $x=\pm it$. This is at $r=0$, $s=\{0,\frac{1}{2}\}$
in our coordinates above, so we expect that they appear simply from
T-dualising the O4 on $x^6$ (now with no NS5 branes complicating the
discussion). This implies the two fixed points have the same NSNS
charge, with an associated projection \emph{opposite} to that of the
gauge factor being confined. Explicitly, this means that if we have a
confining $\Sp$ group, we end up with two fixed points of type O3$^-$
(with one or both possibly of type $\widetilde{\text{O3}^-}$,
depending on discrete gauge and RR flux choices). And similarly, if
the $SO$ group confines we get two orientifolds of type O3$^+$. We
have depicted the confining process in the type IIA picture in
figure~\ref{fig:IIA-confinement}.

It may be illuminating to describe more explicitly the fate of the
deformation $S^3$ after T-duality. The manifold wrapped by the NS5
branes, given by $zw=t^2$, has the topology of a smooth $\bC^*$ when
$t\neq 0$. There is a minimal area $S^1$ in this $\bC^*$, which bounds
a minimal area disk. T-dualising the $x^6$ coordinate over this disk
produces in the dual a $S^1$ fibration over a disk where the fibre
degenerates at the $S^1$ boundary of the disk, a well known
construction of $S^3$.

Let us describe this construction in some detail. We introduce (as we will do
in~\eqref{eq:zi-definition} below) the new coordinates
\begin{equation}
  z_i = (z+w, -i(x+y), x-y, i(w-z))\, .
\end{equation}
In these variables the deformed conifold equation can be written as
\begin{equation}
  \label{eq:IIA-zi-deformed-conifold}
  \sum_{i=1}^4 z_i^2 = 4t^2\, .
\end{equation}
We also have $z_1^2+z_4^2=4zw$, so in these variables the NS5 brane in
the IIA side is wrapping $z_1^2+z_4^2=\frac{1}{4}t^2$. We will
identify below the $S^3$ on the type IIB side as living at
$z_i\in\bR$. This naturally defines a disk
$\Re(z_1)^2+\Re(z_4)^2=\rho^2$, with $\rho\in[0,\frac{1}{2}t]$, ending
on the NS5. T-duality acts as $(z_1,z_4)\to (-z_1,-z_4)$, so any fixed
points must be at the origin of the disk. We expect that in the type IIB
picture the fibre over the origin of the disk is the $S^1$
parametrised by $s$; we will now verify this. From
\eqref{eq:IIA-zi-deformed-conifold}, we have that at the origin of the
disk the $S^1$ fibre in the $S^3$ is given by
$\Re(z_2)^2+\Re(z_3)^2=4t^2$. The locus $r=0$ corresponds to
$(x,y)=(ye^{4\pi is},y)$. We require $xy=-t^2$, so
$y^2e^{4\pi is}=-t^2$, or alternatively $y=ite^{-2\pi is}$,
$x=ite^{2\pi i s}$. Then
\begin{equation}
  (\Re(z_2), \Re(z_3)) = (\Im(x+y), \Re(x-y)) = (2t\cos(2\pi s), -2t\sin(2\pi s))
\end{equation}
precisely according to expectations. So in this notation we see very
clearly that the two O3-planes at $s=\{0,\frac{1}{2}\}$ (equivalently,
at $\Re(z_1)=\Re(z_3)=\Re(z_4)=0$) arise from T-duality of the O4
wrapping the circle T-dual to the $s$ direction, which implies that
they are of the same sign (since in the deformed configuration the NS5
branes do not intersect the O4, so its NSNS charge is the same all
along the circle).

\subsection{The singular orientifolded conifold in type IIB}
\label{sec:singular-orientifolded-conifold}

We will now reproduce and extend these results directly from the type IIB
perspective. There are a number of initially puzzling aspects of the
construction when reinterpreted in this context, as we now discuss. We
will be using the description of fractional branes as coherent
sheaves (see~\cite{Aspinwall:2002ke} for a review which also discusses
the conifold explicitly).

\subsubsection*{Fractional branes and resolved phase}

A useful operation from the IIB perspective is the blow-up of the
singularity. Geometrically, we can think of the singular conifold as a
limit of the blown-up conifold, given by the total space of the
$\cO(-1)\oplus\cO(-1)$ bundle over $\bP^1$. The conifold singularity appears when
the geometric size of the $\bP^1$ goes to zero. In addition to the
geometric volume of the $\bP^1$ one should also consider the integral
of the $B$ field over the $\bP^1$. We have identified the geometric
result of introducing a $B$ field in the T-dual picture in our discussion
above: it corresponds to the relative separation of the two NS5 branes
along the fibre $S^1$. We now would like to identify the effect of
geometrically blowing up the $\bP^1$.

There is basically a unique choice, suggested by analyticity: recall
that the homolorphic K\"ahler coordinate at low energies is $B+iJ$,
with $J$ the volume of the $\bP^1$. Geometrically, the complex
coordinate in the T-dualised conifold is given by $x^6+ix^7$. So, by
holomorphicity, we should identify blow-ups of the $\bP^1$ in the conifold
with displacements of the NS5 branes on the $x^7$ direction. That this
is the right identification can be verified in a number of ways, see
for example \cite{Giveon:1998sr,Kutasov:2012rv}.

The complexified K\"ahler moduli space of the conifold can be
compactified to a $\bP^1$. Let us parametrise this $\bP^1$ of
K\"ahler moduli by a coordinate $\zeta$, with $\zeta=0$ the infinitely
blown-up conifold, and $\zeta=\infty$ the infinitely blown-up conifold
in the flopped phase. The ordinary $B$ and $J$ K\"ahler moduli then
appear as
\begin{align}
  t\equiv B+iJ = \frac{1}{2\pi i}\log \zeta\, .
\end{align}

The two fractional branes in which a D3 decomposes in the conifold
locus can be described in terms of the derived category of coherent
sheaves on the resolved conifold $\widetilde{X}$ (choosing a phase) by
$\cO_C(-1)[1]$ and $\cO_C$, with $C$ the resolution $\bP^1$. The
central charges are fairly easy to compute in this geometry, since
they are uncorrected by world-sheet instantons. They are given by the
large volume expression
\begin{align}
  Z(\cO_C(m)[k]) = (-1)^k (-t + m + 1)\, .
\end{align}
We see that the quiver locus, where the central charges of both
fractional branes are real, is precisely when $t\in\bR$, i.e. $J=0$,
as one may have expected. When in addition $B=0$, one finds that some
of the fractional branes become massless (mass being given by $|Z|$),
so this is a point where light strings can arise. In the type IIA
description this corresponds to the locus in moduli space where the
$x^6$ position of the two NS5 branes coincide.

The type IIA orientifold of interest to us must have a number of surprising
features when reinterpreted in the original language of type IIB at
singularities. First, notice from the IIA description that the
orientifold fixes the NS5 branes to be at $x^7=0$, while allowing
motions in the $x^6$ direction. In IIB language, this can be
reinterpreted as the statement that the orientifold projects out the
size of the resolution $\bP^1$, while preserving the integral of the
$B$ field as a dynamical field. The same point can be seen already
from field theory: the theory with $SO\times \Sp$ group does not admit
Fayet-Iliopoulos terms (simply because there are no $U(1)$s), so there
is no baryonic direction in moduli space. Geometrically, such a
baryonic direction would come from blowing up the singularity: this
would force misalignment between the fractional branes, since they
have opposite BPS phases at large volume. So we also conclude from
this perspective that the blow-up mode must be projected out. This is
somewhat surprising, and contrary to the usual behavior of ordinary
$O3$/$O7$ planes in type~IIB.

A more surprising property (but, as we will see, related to the
previous point) comes again from the fact that the fractional branes
at the conifold admit a description as wrapped D5 and anti-D5
branes. The orientifold that we want must map these fractional branes
to themselves, while being compatible with the supersymmetry preserved
by a background D3. So at the level of the worldsheet it should act as
$\Omega (-1)^{F_L}$, while at the same time somehow mapping a
fractional D3, which is microscopically a wrapped D5, to itself. Our
first goal will be to resolve these tensions.

These issues could be resolved if we take an involution of the
resolved $\bP^1$ that reverses its orientation, such as the $\bZ_2$
action defining the $\bP^1\to \bR\bP^2$ map. Under this involution the
Fubini-Study metric changes sign. So the combined action of $\Omega$
and the geometric action preserves $B$, but not $J$. And similarly,
the D5 wrapping the $\bP^1$ maps to minus itself, which allows it to
survive when combined with the intrinsic minus sign coming from
$\Omega(-1)^{F_L}$. An ordinary D3 is pointlike, so it also
survives. We now show that we do indeed have an orientation reversing
involution.

\subsubsection*{Orientifold geometric involution in the resolved phase}

Recall that the geometric action for our orientifold is given by
\begin{align}
  \label{eq:mixed-involution}
  (x,y,z,w) \to (y,x,-z,-w)\, .
\end{align}
It will be useful to rewrite this action in terms of GLSM fields. The
conifold is described by a GLSM with fields $(x_1,x_2,y_1,y_2)$ with
charges $(1,1,-1,-1)$ under a $U(1)$ gauge group. We take the FI term
to be according to
\begin{align}
  V_D = \left(|x_1|^2 + |x_2|^2 - |y_1|^2 - |y_2|^2 - \xi \right)^2
\end{align}
and the map to the gauge invariant coordinates to be
\begin{align}
  (x,y,z,w) = (x_1y_2, x_2y_1, x_1y_1, x_2y_2)\, .
\end{align}
In these coordinates, the action~\eqref{eq:mixed-involution} is
described by
\begin{align}
  \label{eq:GLSM-involution}
  \sigma\colon (x_1,x_2,y_1,y_2) \to (-y_1, y_2, x_1, -x_2)\, .
\end{align}
There are various things to note in this expression. First, it is a
well defined $\bZ_2$ action when we take the $U(1)$ gauge symmetry into
account: orbits are mapped to orbits. (Even if $\sigma^2=-1$.)

The D-term changes sign, though: if we have a point satisfying the
D-term with $\xi>0$, it will get mapped to a point satisfying the
D-term with $\xi<0$. In other words, the $\bZ_2$ action defines an
involution of the conifold only for the singular conifold, with
$\xi=0$. If $\xi\neq 0$, so we are in some resolved phase, the $\bZ_2$
action maps to the flopped phase: $\xi\to -\xi$. Since $\xi$ can be
interpreted as the volume of the resolved $\bP^1$, this action
achieves precisely what we expected from the general arguments above:
$J=0$ but $B$ is arbitrary, since the volume form in $\bP^1$
geometrically changes sign. In the algebraic language, the statement
is that the $\bZ_2$ acts on the Stanley-Reisner ideal: it exchanges
the Stanley-Reisner ideal $\vev{x_1x_2}$ of a resolved phase ($\xi>0$)
with the Stanley-Reisner ideal $\vev{y_1y_2}$ of the flopped phase
($\xi<0$).

\medskip

For later purposes it will also be useful to describe in more detail
the action of the orientifold on the geometry, which will also give an
explicit proof of the inversion of the volume element of the
resolution $\bP^1$. In particular, we will now describe how the
involution~\eqref{eq:GLSM-involution} acts on the conifold seen as the
real cone over $S^2\times S^3$. We start by reviewing how to go from
the GLSM description in terms of the $x_i,y_i$ variables to the
description in terms of a real cone over $S^2\times S^3$. (The
following discussion of the unorientifolded geometry summarises
\cite{Evslin:2007ux,Evslin:2008ve}, although we deviate slightly from
the presentation there in order to highlight some aspects of the
construction that will become useful to us later.) We will do the
calculation for the singular conifold $\xi=0$. The horizon
$S^2\times S^3$ at a radial distance $r$ is obtained by imposing
\begin{equation}
  |x_1|^2 + |x_2|^2 = |y_1|^2 + |y_2|^2 = r\, .
\end{equation}
We will work at $r=1$ for simplicity. Start by introducing the matrices
\begin{align}
  \begin{split}
    U & = \begin{pmatrix}
      x_1 & -\ov{x_2}\\
      x_2 & \ov{x_1}
    \end{pmatrix}\, ,\\
    V & = \begin{pmatrix}
      \ov{y_1} & -y_2\\
      \ov{y_2} & y_1
    \end{pmatrix}\, .
  \end{split}
\end{align}
It is a simple calculation to show that on the horizon these two
matrices belong to $SU(2)$. Under the $U(1)$ action of the GLSM they
transform as $U \to Ue^{i\alpha\sigma_3}$,
$V\to Ve^{i\alpha\sigma_3}$, with
$\sigma_3=\left(\begin{smallmatrix}1&0\\0&-1\end{smallmatrix}\right)$
the third Pauli matrix. Introduce now the gauge invariants
\begin{align}
  \label{eq:XY-from-UV}
  \begin{split}
    X & = UV^\dagger\,,\\
    Y & = -i U\sigma_3 V^\dagger \,.
  \end{split}
\end{align}
These matrices also clearly belong to $SU(2)$. Following
\cite{Evslin:2007ux}, we also introduce
\begin{align}
  Q = X^\dagger Y = -i V\sigma_3 V^\dagger\, ,
\end{align}
which is nothing but the Hopf projection of $V\in SU(2)=S^3\to
S^2$. It is clear from the second expression that in addition to being
an element of $SU(2)$, $Q$ is traceless, anti-hermitian, and squares
to $-1$. One can also easily see that there is a bijection between the
pair $(X,Q)$ and the usual set of coordinates for the conifold
\begin{align}
  W = \begin{pmatrix}
    x_1y_1 & x_1 y_2\\
    x_2y_1 & x_2y_2
  \end{pmatrix} = \frac{1}{2}(X+iY)=\frac{1}{2}X({\bf 1} + iQ)\, .
\end{align}
That the bijection exists is manifest if we construct $X$, $Y$ in
terms of $W$ as follows
\begin{align}
  \begin{split}
    X & = \Tr W^\dagger + W - W^\dagger\\
    Y & = i\Tr W^\dagger - i(W - W^\dagger)\, .
  \end{split}
\end{align}
Now, $X$ and $Q$ are independent, so they parametrise a product
space. $X$ is a generic $SU(2)$ matrix, so it parametrises a $S^3$,
while the condition that $Q$ is a traceless $SU(2)$ matrix implies
that it parametrises an $S^2$. We thus have a good set of coordinates
for $S^2\times S^3$, and we showed explicitly the diffeomorphism to
the conifold base in the usual coordinates. It will be convenient to
be more explicit about the coordinates of the spheres. For a generic
$SU(2)$ matrix $S$ one has the Pauli decomposition
\begin{align}
  \label{eq:Pauli-decomposition}
  S = S_0 + i \sum_{i=1}^3 S_i \sigma_i
\end{align}
with $\sigma_i$ the Pauli matrices, $S_0 = \frac{1}{2}\Tr S$,
$S_i=-\frac{i}{2} \Tr (S\sigma_i)$. $\det S=1$ implies
$\sum_{\mu=0}^3 S_\mu^2=1$, which is the usual equation of
$S^3\subset \bR^4$. Imposing tracelessness of $S$, as for $Q$, sets
$S_0=0$, and thus gives a $S^2\subset S^3$, as we claimed above. In
what follows we denote by $U_\mu,V_\mu,X_\mu,Y_\mu,Q_\mu$ the
components of the $SU(2)$ matrices $U,V,X,Y,Q$ in this basis.

\medskip

With this description of the $S^2\times S^3$ horizon of the conifold
in hand we can come back to the orientifold
action~\eqref{eq:GLSM-involution}. In terms of the projective
coordinates
\begin{equation}
  \begin{split}
    X_\mu & = \biggl(\frac{1}{2} \, x_{1} y_{1} + \frac{1}{2} \, x_{2}
    y_{2} + \frac{1}{2} \, \overline{x_{1}} \overline{y_{1}} +
    \frac{1}{2} \, \overline{x_{2}} \overline{y_{2}}\, ,\\
    & \phantom{= \biggl(} -\frac{1}{2} i \, x_{2} y_{1} - \frac{1}{2} i \, x_{1} y_{2} +
    \frac{1}{2} i \, \overline{x_{2}} \overline{y_{1}} + \frac{1}{2}
    i \, \overline{x_{1}} \overline{y_{2}}\, ,\\
    & \phantom{= \biggl(} -\frac{1}{2} \, x_{2} y_{1} + \frac{1}{2} \, x_{1}
    y_{2} - \frac{1}{2} \, \overline{x_{2}} \overline{y_{1}} +
    \frac{1}{2} \, \overline{x_{1}} \overline{y_{2}}\, ,\\
    & \phantom{= \biggl(} -\frac{1}{2} i \, x_{1} y_{1} + \frac{1}{2} i \,
    x_{2} y_{2} + \frac{1}{2} i \, \overline{x_{1}} \overline{y_{1}}
    - \frac{1}{2} i \, \overline{x_{2}} \overline{y_{2}}\biggr)\, .
  \end{split}
  \label{eq:Xmu}
\end{equation}
We can rewrite this equation in terms of the GLSM invariant
coordinates $x,y,z,w$ as
\begin{equation}
  X_\mu = (\Re(z+w), \Im(x+y), \Re(x-y), \Im(z-w))\, .
\end{equation}
This suggests introducing the new variables
\begin{equation}
  \label{eq:zi-definition}
  z_i=(z+w, -i(x+y),x-y, i(w-z))\, ,
\end{equation}
so that
\begin{equation}
  \label{eq:Xmu-z}
  X_\mu = (\Re (z_1), \Re(z_2), \Re(z_3), \Re(z_4))\, .
\end{equation}
In terms of these variables the conifold equation $xy=zw$ becomes
\begin{equation}
  \label{eq:conifold-z}
  \sum_{i=1}^4 z_i^2 = 0 \, ,
\end{equation}
and the deformed conifold equation $zw=xy+t^2$ becomes
\begin{equation}
  \label{eq:zi-deformed-conifold}
  \sum_{i=1}^4 z_i^2 = 4t^2\, .
\end{equation}
The involution~\eqref{eq:mixed-involution} acts on these variables as
\begin{equation}
  \label{eq:z-involution}
  z_i \to (-z_1, z_2, -z_3, -z_4)\, .
\end{equation}
From here, or directly doing a bit of algebra on~\eqref{eq:Xmu}, one
finds that the action~\eqref{eq:GLSM-involution} on the $S^3$
coordinates $X$ is given by
\begin{equation}
  X_\mu \to (-X_0, X_1, -X_2, -X_3)
\end{equation}
which has fixed points (forgetting about the $S^2$ momentarily) at
$X_0=X_2=X_3=0$, i.e. two points in the $S^3$. This agrees with the
fixed point structure we found from our type IIA picture 
in~\S\ref{sec:IIA-interpretation}.

Let us study the structure of the $S^2$ component at one of these
fixed points in the $S^3$. Going to the patch $x_1\neq 0$ we can gauge
fix $x_1$ to be real and positive. A solution to $X_0=X_2=X_3=0$ can
then be found at $(x_1,x_2,y_1,y_2)=(1,0,0,i)$, which maps to
$X_\mu=(0,1,0,0)$. As a small consistency check, notice that the
action of~\eqref{eq:GLSM-involution} on this point gives
$(x_1,x_2,y_1,y_2)=(0,i,1,0)$, which again maps to $X_\mu=(0,1,0,0)$,
but as expected acts freely on the total space $S^3\times S^2$. To
reconstruct the whole $\bP^1$ we start with the point
$(x_1,x_2,y_1,y_2)=(1,0,0,i)$, giving
\begin{equation}
  U_0 = \begin{pmatrix}
    1 & 0\\
    0 & 1
  \end{pmatrix} = \sigma_0 \qquad ; \qquad V_0 = \begin{pmatrix}
    0 & -i\\
    -i & 0
  \end{pmatrix} = -i\sigma_1\, .
\end{equation}
Tracing through the definitions, this gives $X=i\sigma_1$,
$Y=i\sigma_2$ and $Q=i\sigma_3$. Any other point in the $\bP^1$ above
$(U_0,V_0)$ can be written as $(U,V)=(U_0g, V_0g)$ for some
$g\in SU(2)$. This leaves $X=U_0V_0^\dagger$ invariant, but introduces
a dependence of $Q=-i \sigma_1g\sigma_3 g^{-1}\sigma_1$ on $g$.

In terms of $U,V$ the action~\eqref{eq:GLSM-involution} acts as
\begin{equation}
  U \to -\sigma_1 V \sigma_1 \qquad ; \qquad V \to \sigma_1 U
  \sigma_1
\end{equation}
so it sends
\begin{equation}
  Q=-iV\sigma_3V^\dagger \to -i \sigma_1 U \sigma_1 \sigma_3 \sigma_1
  U^\dagger \sigma_1 = i\sigma_1 U \sigma_3 U^\dagger \sigma_1
\end{equation}
which for the $\bP^1$ we are studying reduces to
\begin{equation}
  Q_g= -i\sigma_1
  g\sigma_3 g^{-1} \sigma_1 \to i\sigma_1 g \sigma_3 g^{-1}\sigma_1 =
  -Q_g\, .
\end{equation}
So we learn that the action of the involution on the $\bP^1$ above
$(U_0, V_0)$ is the orientation-reversing $\bP^1\to \bR\bP^2$ map, as
we guessed above based on the IIA dual and microscopic
considerations. There is also a second fixed point at
$X_\mu=(0,-1,0,0)$, for which a very similar discussion applies.

\subsection{The orientifolded cascade}

The discussion in the previous section was about the singular
conifold. In analogy with the behavior in absence of the orientifold
\cite{Klebanov:2000hb}, for nontrivial fractional brane configurations
the orientifolded conifold is deformed dynamically. In this section,
we want to study this effect from the field theory point of view. In
particular, by this method we will verify the prediction for the
orientifold charges given in \S\ref{sec:IIA-interpretation}. Useful
references for this section are
\cite{Dymarsky:2005xt,Krishnan:2008gx}.

\subsubsection*{Classical dynamics}

The dimer model and the quiver describing the low energy dynamics for
D3-branes on the orientifold of the conifold we are studying were
given in figure~\ref{fig:orientifolded-conifold}. The superpotential
for the resulting theory is somewhat subtle, but its form is important
for the considerations below, so let us derive it in some detail. We
parametrise the fields of the $SU(N)\times SU(M)$ theory before
taking the orientifold as $A_i,B_i$, with
$A_i\in(\fund_M, \ov{\fund}_N)$ and $B_i\in(\ov\fund_M, \fund_N)$.
The superpotential for this theory is well known
\cite{Klebanov:1998hh}:
\begin{equation}
  \label{eq:conifold-W}
  W = \frac{1}{2}\varepsilon_{ij}\varepsilon_{lm}\Tr\left(A_iB_lA_jB_m\right) = \Tr\left(A_1B_1A_2B_2 - B_2A_2B_1A_1\right)\, .
\end{equation}
There is a $SO(4)=SU(2)_1\times SU(2)_2$ global symmetry of the
singular conifold. In terms of the GLSM it manifests itself as
$(x_i,y_i)\to (g_1\bm{x}, g_2\bm{y})$, with $g_i\in SU(2)_i$ in the
fundamental representation, and $\bm{x}=(x_1,x_2)$,
$\bm{y}=(y_1,y_2)$. For the case $N=M=1$ of a single brane probing the
conifold we can identify $\vev{A_i}=x_i$, $\vev{B_i}=y_i$. The
involution~\eqref{eq:GLSM-involution} can be written in these
variables as
\begin{equation}
  \sigma\colon (\bm{x},\bm{y}) \to (-\sigma_3\bm{y}, \sigma_3\bm{x})\, .
\end{equation}
We want to determine the subgroup $G\subset SU(2)_1\times SU(2)_2$
compatible with $\sigma$. That is, for every $g\in G$,
$\sigma g = g\sigma$, modulo the $U(1)$ GLSM action
$(\bm{x},\bm{y})\to(e^{i\alpha}\bm{x},
e^{-i\alpha}\bm{y})$.
Equivalently, in block matrix form
\begin{equation}
  \begin{pmatrix}
    g_1 & 0 \\
    0 & g_2
  \end{pmatrix} =
  -\begin{pmatrix}
    0 & -\sigma_3\\
    \sigma_3 & 0
  \end{pmatrix}
  \begin{pmatrix}
    g_1 & 0\\
    0 & g_2
  \end{pmatrix}
  \begin{pmatrix}
    0 & -\sigma_3\\
    \sigma_3 & 0
  \end{pmatrix} =
  \begin{pmatrix}
    \sigma_3 g_2\sigma_3 & 0 \\
    0 & \sigma_3 g_1 \sigma_3
  \end{pmatrix}
\end{equation}
which can be satisfied by $g_1=\sigma_3g_2\sigma_3$. Parametrising
$g_1= a_0+i\sum_{k=1}^3 a_k\sigma_k$,
$g_2=b_0+i\sum_{k=1}^3 b_k\sigma_k$ (with
$\sum a_\mu^2=\sum b_\mu^2=1$), this requires
$(a_0,a_1,a_2,a_3)=(b_0,-b_1,-b_2,b_3)$. So we learn that
$SU(2)_d\subset G$ is conserved by the orientifold action.

Let us come back to the field theory arising after orientifolding,
described by the quiver in
figure~\ref{sfig:orientifolded-conifold-quiver}. From the action of
the involution on the dimer model in
figure~\ref{sfig:orientifolded-conifold-dimer} we immediately read
that the invariant fields under the involution satisfy
\begin{equation}
  \label{eq:orientifold-fields}
  \begin{split}
    B_1 & = s_1 \gamma_{\Sp} A_1^t \gamma_{SO}\, ,\\
    B_2 & = s_2 \gamma_{\Sp} A_2^t \gamma_{SO}\, .
  \end{split}
\end{equation}
We take the following block-diagonal representation for the Chan-Paton
matrices
\begin{equation}
  \label{eq:gammas}
  \gamma_{SO} = \bm{1} \qquad ; \qquad \gamma_{\Sp} = \begin{pmatrix}
    \sigma_2 \\
    & \sigma_2 \\
    & & \ddots \\
    & & & \sigma_2
  \end{pmatrix}
\end{equation}
with
$\sigma_2=\left(\begin{smallmatrix}0 & -i\\ i &
    0\end{smallmatrix}\right)$
for the action of the orientifold on the gauge factors. The transpose
in~\eqref{eq:orientifold-fields} is, as usual, coming from the
reflection of the worldsheet. We have additionally included a possible
sign $s_i=\pm 1$ for completeness. We can nevertheless now use our
observation of the presence of the $SU(2)_d$ symmetry after
orientifolding to impose $s_1=s_2$, and then redefine these signs
away. We will set $s_i=+1$ in what follows.

We thus find that the superpotential after orientifolding is
\begin{equation}
  \label{eq:orientifolded-W}
  \begin{split}
    W & = \frac{1}{4}
    \varepsilon_{ij}\varepsilon_{lm}\Tr\left(A_i\gamma_{\Sp}A_l^t\gamma_{SO}A_j\gamma_{\Sp}A_m^t\gamma_{SO}\right)\, .
  \end{split}
\end{equation}
As one may have guessed, this is the projection of the original
superpotential to the invariant fields, and it preserves the $SU(2)$
symmetry we have identified geometrically above.

\medskip

Let us try to gain some intuition for this theory, before we start
analysing the cascade. A simple thing to try is to construct the
classical moduli space of a single mobile brane probing the
geometry. (The following analysis was also done in~\cite{Imai:2001cq},
but the details of the argument will be slightly different since our
convention~\eqref{eq:gammas} for the Chan-Paton matrices is different,
so we include it here since it may be illuminating for later
discussion.)

When the brane is at the singularity, the gauge algebra is
$\fso(2)\oplus \fsp(2)$. (The gauge group has in addition a gauged
$\bZ_2$ external automorphism, and is more precisely
$O(2)\times \Sp(2)$.) In this case we can treat the $A_i$ fields as
$2\times 2$ matrices, transforming under $(g,h)\in O(2)\times\Sp(2)$
as $A_k \to g A_k h$. The non-abelian D-terms for $\fso(2)$ are
\begin{equation}
  \label{eq:SO(2)-D-term}
  \sum_{k=1}^2\Tr(A_k^\dagger\sigma_2A_k) = 0 \:,
\end{equation}
while the non-abelian D-terms for $\fsp(2)$ are
\begin{equation}
  \label{eq:Sp(2)-D-terms}
  \sum_{k=1}^2\Tr(A_k\sigma_i A_k^\dagger) = 0
\end{equation}
for any Pauli matrix $\sigma_i$.

A generic solution of the F-term coming
from~\eqref{eq:orientifolded-W}, together with the
D-terms~\eqref{eq:SO(2)-D-term} and~\eqref{eq:Sp(2)-D-terms} can be
written as
\begin{equation}
  \label{eq:SO(2)xSp(2)-ansatz}
  A_1 = \begin{pmatrix}
    \sx_1 & \sy_1\\
    i\sx_1 & -i\sy_1
  \end{pmatrix} \,\qquad \qquad
  A_2 = \begin{pmatrix}
    \sx_2 & \sy_2\\
    i\sx_2 & -i\sy_2
  \end{pmatrix} \, ,
\end{equation}
subject to the condition
\begin{equation}
  \label{eq:orientifolded-NAD}
  |\sx_1|^2 + |\sx_2|^2 - |\sy_1|^2 - |\sy_2|^2 = 0\, .
\end{equation}
We still have a remnant of the $O(2)\times \Sp(2)$ symmetry acting on
$A_i$, while keeping the form~\eqref{eq:SO(2)xSp(2)-ansatz}. These are
$\Sp(2)$ transformations acting as
\begin{equation}
  \label{eq:Sp(2)-remaining-U(1)}
  A_1 \to A_1\begin{pmatrix}
    e^{i\alpha} & 0\\
    0 & e^{-i\alpha}
    \end{pmatrix}
\end{equation}
which in terms of the $\sx_i,\sy_i$ components is
$(\sx_1,\sx_2,\sy_1,\sy_2)\to (e^{i\alpha}\sx_1, e^{i\alpha}\sx_2,
e^{-i\alpha}\sy_1, e^{-i\alpha}\sy_2)$.
This, together with the D-term~\eqref{eq:orientifolded-NAD},
reproduces the standard GLSM construction for the singular
conifold. In addition, we have the external $\bZ_2$ automorphism,
which acts as $A_i\to \sigma_3 A_i$. Combining this action with an
appropriate $\Sp(2)$ transformation we obtain an extra $\bZ_2$ action
leaving the form of the solution~\eqref{eq:SO(2)xSp(2)-ansatz}
invariant
\begin{equation}
  A_i \to \sigma_3 A_i (i\sigma_2) = \begin{pmatrix}
    -\sy_i & \sx_i\\
    -i\sy_i & -i\sx_i
  \end{pmatrix}
\end{equation}
or directly in terms of the GLSM coordinates $(\sx_i,\sy_i)\to
(-\sy_i, \sx_i)$, which perfectly reproduces~\eqref{eq:GLSM-involution}
(up to a harmless sign redefinition).

\subsubsection*{Quantum dynamics}

Now that we have an understanding of the single probe brane case in
the classical setting, let us move on to the calculation of interest,
namely the determination of the properties of the mesonic branch of
the deformed orientifolded conifold, when we have more than one mobile
brane probing the dynamics. (We take more than one brane in order to
be able to more clearly study $O$ and $\Sp$ enhancements at the
conifold loci.)  The case without the orientifold has been extensively
studied, some useful references are
\cite{Klebanov:2000hb,Dymarsky:2005xt,Krishnan:2008gx}. The
orientifolded case has been studied (in part, we will need to extend
the analysis) in \cite{Imai:2001cq}. A first easy observation is that
the seem to be various basic channels for confinement in the
$O(Q)\times \Sp(2P)$ theory. If $Q\gg P$ the $O(Q)$ node will confine
first, and we will end up with a theory of $\Sp(2P)$ adjoint
mesons. Similarly, if $P\gg Q$ confinement in the $\Sp(2P)$ node will
occur first, so we will have a theory of $O(Q)$ adjoint mesons.

We want to understand the nature of the O3-planes after confinement in
each of these cases. From the IIA perspective we expect that when
$\Sp(2P)$ confinement dominates we end up with O3$^-$ planes. In the case
where $Q\in 2\bZ$ we expect the two O3$^-$ planes to be of the same
type (either both O3$^-$ or both $\tO^-$), while in the $Q\in 2\bZ+1$
we expect one O3$^-$ and one $\tO^-$. In the case where the $O(Q)$
node confines first we expect the two O3-planes to be O3$^+$. In this
case we cannot say whether they are O3$^+$ or $\tO^+$ with the
techniques in this section, since they lead to identical perturbative
physics, but this distinction is not interesting for our model
building purposes in any case.

We will focus on $\Sp(2P)$ confinement driving the
dynamics.\footnote{The first part of the analysis in this section can
  already be found (in slightly different conventions) in the
  literature \cite{Ahn:2001hy,Imai:2001cq}, but we include it both for
  completeness, and to motivate the later part of the discussion,
  where we study the enhanced symmetry loci in the moduli space in
  order to probe the nature of the resulting orientifold fixed points
  after confinement. The result we find agrees with the expectations
  from the type IIA picture (and thus disagrees with the results
  claimed in \cite{Ahn:2001hy,Imai:2001cq}).} In order to have a
weakly coupled geometry after confinement we require $P\gg 1$.  In
this case we expect to end up with two O3$^-$ planes of the same or
different type, depending on the parity of $Q$. We choose to analyze
$Q\in 2\bZ$, since it makes the analysis a little bit simpler, and is
seems to be the most convenient one for model building purposes: the
D3 charge of the orientifold system is integral, as opposed to
half-integral. The rest of the cases can be analysed very similarly,
confirming the IIA predictions just mentioned, so we omit their
explicit discussion.

\medskip

The confined description is in terms of gauge invariant mesons
\begin{equation}
  \cM_{ij} = A_i\gamma_{\Sp}A_j^t\, .
\end{equation}
In order to understand the dynamics of the probe stack, consider again
the classical moduli space of a stack of $k$ mobile D3-branes. We
can construct it by choosing block-diagonal and equal vevs for the
$2k\times (2P+2k)$ matrix $A_i$
\begin{equation}
  \label{eq:Sp-ansatz}
  A_i = \begin{pmatrix}
    \bm{\sx}_i & & & & \bm{0}_{k,2P}\\
     & \bm{\sx}_i & & & \bm{0}_{k,2P}\\
     & & \ddots & & \vdots\\
     & & & \bm{\sx}_i & \bm{0}_{k,2P}
  \end{pmatrix}
\end{equation}
with $\bm{0}_{r,s}$ is the $r\times s$ zero matrix, and
\begin{equation}
  \bm{\sx}_i = \begin{pmatrix}
    \sx_i & \sy_i \\
    i\sx_i & -i\sy_i
  \end{pmatrix}
\end{equation}
as in~\eqref{eq:SO(2)xSp(2)-ansatz}. The classical $\Sp$ mesons,
transforming in the adjoint of $O(2k)$, are given by
\begin{equation}
  \label{eq:Sp-M}
  \cM_{ij} = \underbrace{\begin{pmatrix}
     \bm{\sx}_i\sigma_2 \bm{\sx}_j^t \\
     & \bm{\sx}_i\sigma_2 \bm{\sx}_j^t\\
     && \ddots \\
     &&& \bm{\sx}_i\sigma_2 \bm{\sx}_j^t
   \end{pmatrix}}_{k\text{ blocks}}
   \, .
\end{equation}

In the confined description the mesons become elementary fields. The
classical picture suggests parametrising the moduli space of mesons
in the following way. Introduce the basic elementary meson $\sz_{ij}$,
which in the classical limit can be written as
\begin{equation}
  \sz = \begin{pmatrix}
    \sx_1\sy_1 & \sx_1\sy_2\\
    \sx_2\sy_1 & \sx_2\sy_2
  \end{pmatrix}\, .
\end{equation}
We parametrise the possible space of vacua by rewriting the classical
expressions for the mesons in~\eqref{eq:Sp-M} by their expression in
terms of the fundamental mesons $\sz_{ij}$:
\begin{equation}
  \label{eq:Sp-M-ansatz}
  \cM_{ij} = \begin{pmatrix}
    \bsm_{ij} \\
    & \ddots \\
    && \bsm_{ij}
  \end{pmatrix}
\end{equation}
with
\begin{equation}
  \label{eq:Sp-m-ansatz}
  \begin{split}
    \bsm_{11} = \begin{pmatrix}
      0 & -2\sz_{11}\\
      2\sz_{11} & 0
    \end{pmatrix} \quad & ; \quad
    \bsm_{12} = \begin{pmatrix}
      i(\sz_{21} - \sz_{12}) & -(\sz_{12}+\sz_{21})\\
      \sz_{12} + \sz_{21} & i(\sz_{21} - \sz_{12})
    \end{pmatrix}\\
    \bsm_{21} = \begin{pmatrix}
      i(\sz_{12} - \sz_{21}) & -(\sz_{12}+\sz_{21})\\
      \sz_{12} + \sz_{21} & i(\sz_{12} - \sz_{21})
    \end{pmatrix} \quad & ; \quad
    \bsm_{22} = \begin{pmatrix}
      0 & -2\sz_{22} \\
      2\sz_{22} & 0
    \end{pmatrix}
  \end{split}
\end{equation}

\medskip

One can easily see that these vevs satisfy the non-abelian D-term
conditions for $O(2k)$. The F-terms are satisfied as follows. In the
confined mesonic variables the classical
superpotential~\eqref{eq:orientifolded-W} becomes
\begin{equation}
  W = \frac{1}{4}\varepsilon_{ij}\varepsilon_{lm}\Tr\left(
    \cM_{il} \gamma_{SO} \cM_{jm} \gamma_{SO}
  \right)\, .
\end{equation}
It is well know that this superpotential gets modified
non-perturbatively to \cite{Terning:2003th}
\begin{equation}
  \label{eq:Sp-W}
  W = \frac{1}{4}\varepsilon_{ij}\varepsilon_{lm}\Tr\left(
    \cM_{il} \gamma_{SO} \cM_{jm} \gamma_{SO}
  \right) + \left(\frac{\Lambda^{\frac{b}{2}}}{\Pf([\cM])}\right)^{\frac{1}{P-k+1}}
\end{equation}
with $\Lambda$ the dynamical scale of the $\Sp$ node, $b=2(3P+k+3)$
the one-loop $\beta$ function coefficient of the $\Sp$ theory, and
\begin{equation}
  [\cM] = \begin{pmatrix}
    \cM_{11} & \cM_{12} \\
    \cM_{21} & \cM_{22}
  \end{pmatrix}\, .
\end{equation}

The F-term equations then imply for the
ansatz~\eqref{eq:Sp-M-ansatz} that
\begin{equation}
  \label{eq:Sp-field-theory-conifold}
  \det(\sz) = \Lambda^{\frac{b}{2(P+1)}}
\end{equation}
ignoring some irrelevant numerical constants. This is precisely the
equation for the deformed conifold, with the small subtlety of the
presence of a branch structure (due to the $2(P+1)$-th root),
associated with the flux appearing after confinement
\cite{Klebanov:2000hb}.

\medskip

In order to determine the nature of the orientifolds we need to
determine the subgroup of $O(2k)$ leaving invariant all the meson
vevs~\eqref{eq:Sp-M} for all points in the moduli space. It is not
hard to see that at generic points in moduli space the preserved gauge
symmetry is $U(k)$. We interpret this as the $U(k)$ theory on the
D3 probe stack away from any enhancement points.

In the current field theory conventions, the orientifold involution
(encoded in the $\bZ_2$ automorphism part of the $O(2k)$ gauge group)
acts on the moduli space as
\begin{equation}
  (z_{11}, z_{12}, z_{21}, z_{22}) \to (-z_{11}, - z_{21}, - z_{12}, -z_{22})
\end{equation}
so there are fixed points of the involution at $z_{11}=z_{22}=0$,
$z_{12}=- z_{21}$.  Notice from~\eqref{eq:Sp-field-theory-conifold}
that there are exactly two such points in the moduli space for each
branch of moduli space, coming from
$\sz_{12}^2=\Lambda^{\frac{b}{2(P+1)}}$. At these two points in moduli
space we have $\cM_{11}=\cM_{22}=0$, and
$\cM_{12}=-\cM_{21}\propto\bm{1}$, so the $O(2k)$ gauge group is
unbroken. The natural interpretation of these points in moduli space
is as the locations where the probe stack of branes comes on top of
the two orientifold planes that we expect. Since both enhancements are
to $O(2k)$, this shows that both orientifold planes are O3$^-$ planes.

\subsection{Orientifold type changing transitions}
\label{sec:orientifold-transition}

There is one small loose end in this whole discussion. Assume that we
do not put any (fractional or regular) branes on the conifold. It
seems like we have a choice in whether we deform into the
configuration with two O3$^-$ or two O3$^+$ planes, and furthermore,
these two configurations seem to be smoothly connected by a local
operation on the conifold. On the other hand, these two configurations
have opposite RR charge, differing in the charge of a mobile D3. This
is measurable asymptotically, so we have a puzzle.

A careful formulation of the puzzle leads almost immediately to the
solution. Notice that, since the O4$^-$ and O4$^+$ planes have
opposite RR charge, in the absence of fractional branes the type IIA
configuration does not have the same tension on both sides of the NS5
branes, and the O4$^+$ side will tend to confine. This may perhaps
sounds surprising, but it is a manifestation of the fact that isolated
$\Sp(0)$ nodes in string theory behave as if there was gaugino
condensation on them
\cite{Intriligator:2003xs,Aganagic:2007py,GarciaEtxebarria:2008iw}. In
order to truly have the two kinds of orientifold configurations
connected in moduli space, we need to balance the tension by adding
two fractional branes on the $SO$ side, giving rise to a
$SO(2)\times \Sp(0)$ theory. In this case the $\Sp(0)$ node no longer
confines, due to the extra flavors.

For the $SO(2)\times \Sp(0)$ theory, where one does have a moduli
space connecting both types of configurations, the contradiction
evaporates: if we deform by contracting the $SO(2)$ side to nothing we
end up with two O3$^+$ planes at the fixed points, while if we deform
by contracting the $\Sp(0)$ side we end up with two O3$^-$ planes and
a mobile D3-brane (or alternatively two $\tO^-$
planes with no D3, depending on which branch of moduli space we
choose), which has the same overall D3 charge.

\subsection{Decay to a supersymmetric configuration}
\label{sec:decay}

The supersymmetry breaking system of interest to us, realising the
nilpotent Goldstino, can now be easily engineered by putting a stuck
D3 on top of one of the O3$^-$ planes, and a stuck $\ad3$ on top of
the other O3$^-$. We emphasize that this is certainly not the only
choice, particularly in the models below where we have more than two
O3-planes, but we find it convenient, since in this way one can add a
nilpotent Goldstino sector to an existing supersymmetric model without
affecting the tadpoles.

If we arrange branes in this way there is an interesting
non-perturbative decay channel, somewhat similar to the one in
\cite{Kachru:2002gs}, that we now discuss briefly.\footnote{Notice
  that in contrast with the decay process in \cite{Kachru:2002gs}, in
  our case we have a single stuck \ad3, so no polarization due to the
  non-abelian interaction with the fluxes \cite{Myers:1999ps} is
  possible. Thus the perturbative decay channel in
  \cite{Kachru:2002gs}, present when the number of \ad3 branes is
  large enough compared to the flux, is always absent in our setting.}
Recall from \cite{Witten:1998xy,Hyakutake:2000mr} that in flat space
the stuck D3 brane on top of the O3$^-$, or in other words the
$\tO^-$, can be alternatively described by a D5-brane wrapping the
topologically nontrivial $\bR\bP^2\in H_2(\bR\bP^5,\widetilde{\bZ})$
around the O3$^-$. This D5 dynamically decays onto the O3$^-$, and
produces the $\tO^-$.

\begin{figure}
  \centering
  \includegraphics[width=\textwidth]{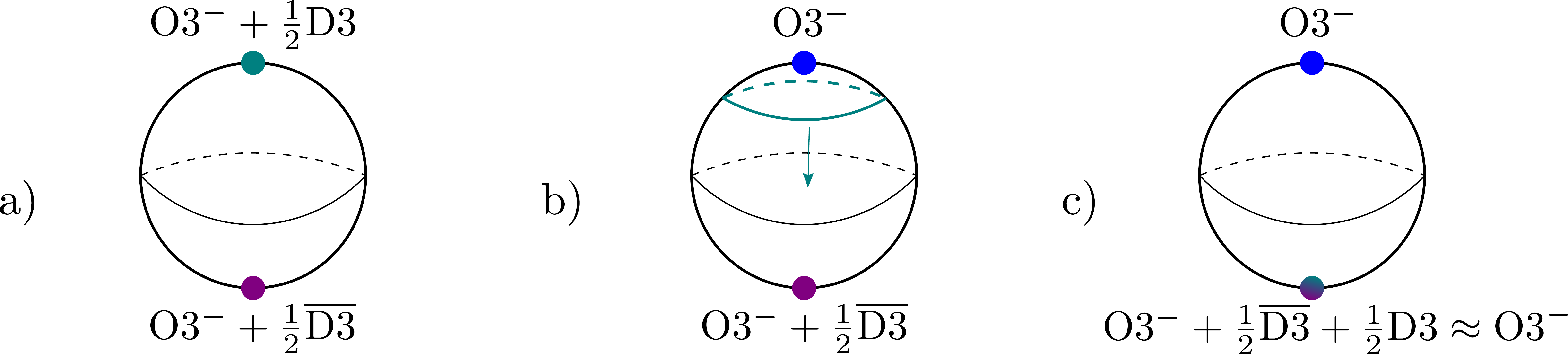}
  \caption{Non-perturbative decay process into the supersymmetric
    configuration. a) The original configuration with a nilpotent
    Goldstino. We display the $S^3/\bZ_2$ at the bottom of the throat.
    b) We resolve the stuck D3-brane into a D5 wrapping
    $\bRP^2\in S^3/\bZ_2$. c) We close the $\bR\bP^2$ over the other
    orientifold fixed point, and tachyon condensation takes over,
    rolling down to the supersymmetric vacuum.}
  \label{fig:decay}
\end{figure}

If we adapt this discussion to the case of the two O3-planes at the
bottom of the cascade with stuck D3 and \ad3 branes, we have that we
can resolve the stuck D3-brane (say) into a D5-brane wrapping the
$\bR\bP^2$ at the equator of the $S^3/\bZ_2$ at the bottom of the
cascade, and then close this D5 on the other side of the $S^3/\bZ_2$,
where the \ad3 is stuck. The resulting system has an ordinary O3$^-$
on one fixed point on the $S^3/\bZ_2$, and a D3-\ad3 pair stuck on top
of the other O3$^-$. The brane-antibrane pair can then annihilate by
ordinary tachyon condensation, and we return to the original
supersymmetric vacuum. We show the process in
figure~\ref{fig:decay}.

\section{Global embeddings}
\label{sec:global-embeddings}

Now that we understand the local dynamics in detail, let us try to
construct a global example exhibiting these dynamics. The conifold
singularity is ubiquitous in the space of Calabi-Yau
compactifications. It is, however, less easy to find a space that
admits the involution described above and allows for the cancellation
of all tadpoles.

We may try to find global embeddings on the ``resolution phase'', or
on the ``deformation phase''. In the first case, we try to construct a
toric space such that is has a conifold singularity admitting the
desired involution. The simplest construction would have the conifold
realised as a face of the toric polytope, as in
\cite{Balasubramanian:2012wd}. These are ubiquitous, as discussed in
that paper. One should then search the subclass of models compatible
with an involution of the form~\eqref{eq:GLSM-involution}.

We can instead choose to search for models in the ``deformation
phase'', namely directly on the side described by flux and a $S^3$
with O3-planes at the north and south poles of a deformed conifold,
which is the description of interest for model building. In this paper
we focus on models directly described in the deformation phase,
leaving the search for models in the resolution phase for future work.

A possible search strategy is as follows: first we construct
consistent type IIB models with O3-planes in them. We choose to construct
them by giving a Calabi-Yau threefold, and specifying an appropriate
involution on it.  Then we try to deform the complex structure such
that two O3-planes are brought together. If this is possible, we
analyze the topology of the resulting Calabi-Yau (before taking the
involution) to see whether the neighborhood of the points where the O3
planes coincide is locally a conifold. If so, we need to verify
whether tadpoles can be cancelled.  This is the way we found one of
our two examples, that we describe in \S\ref{Sec:secondEXAMPLE}.
An exhaustive search may produce several candidates. We leave this for
future work.\footnote{In some cases we might obtain more drastic
  configurations, in which more than two O3-planes come
  together.}

In our first example, however, we present a variation of this
strategy. One can construct some (non-Calabi-Yau) base $\cB$, with
some interesting properties to be discussed momentarily, over which we
fibre a torus in such a way that we obtain a Calabi-Yau fourfold. We
then take the F-theory limit in order to produce the desired type IIB
background. The defining property of $\cB$ is that it has one or more
(deformed) conifold singularities, and that the local
involution~\eqref{eq:mixed-involution} extends to an involution
$\sigma$ over $\cB$. The fourfold of interest will then be a
Calabi-Yau genus one fibration over $\sB=\cB/\sigma$. Over the fixed
points in $\sB$ we will find a local structure of the form
$\bC^3/\bZ_2$, with the $\bZ_2$ acting as $(x,y,z)\to (-x,-y,-z)$,
which is not Calabi-Yau. The Calabi-Yau fourfold will then promote
this local structure to four terminal $\bC^4/\bZ_2$ singularities,
which signal the appearance of an O3.

These conditions on $\cB$ do not seem very restrictive, so we expect
to be able to find examples with relative ease. We again leave a
systematic classification for the future.  Here we present a simple
example of this sort, that we now discuss.

\subsection{F-theory construction}\label{Sec:FirstEXAMPLE}

One early well-known example of conifold embedding in the string
phenomenology literature already exhibits the structure we want
\cite{Giddings:2001yu}. Take $\cB$ to be defined by a quartic on
$\bP^4$ of the form
\begin{equation}
  \label{eq:B-hypersurface}
  P = \sum_{i=1}^4 (z_5^2 + z_i^2) z_i^2 - t^2z_5^4 = 0\, .
\end{equation}
We have intentionally abused notation, and denoted the projective
coordinates of $\bP^4$ by $z_1,\ldots,z_5$, similarly to a set of
coordinates used for the conifold above. As in \cite{Giddings:2001yu},
we choose $t$ to be real. It is then easily checked that $\cB$ is
smooth for $t\neq 0$, but develops a conifold singularity at
$[0:0:0:0:1]$ when $t=0$. In a neighborhood of this point
we can gauge-fix $z_5=1$, obtaining a local structure
\begin{equation}\label{eq:localPnoblp}
  \widetilde{P} = \sum_{i=1}^4 (1 + z_i^2)z_i^2 - t^2 = 0
\end{equation}
which for small enough $z_i$ is the standard form of the conifold.

The involution~\eqref{eq:z-involution} is then clearly extensible to
$\cB$, by taking
\begin{equation}
  \label{eq:sigma-involution}
  \sigma\colon (z_1,z_2,z_3,z_4,z_5) \to (-z_1, z_2, -z_3, -z_4, z_5)\, .
\end{equation}
The fixed loci are at $z_1=z_3=z_4=0$ and $z_2=z_5=0$. (We will later 
blow-up the latter). The first set consists of four points in
$\cB$, given by the solutions of $(z_5^2+z_2^2)z_2^2 - t^2z_5^4$. In
all four points we need to have $z_5\neq 0$ (otherwise
$z_1=\ldots=z_5=0$, which is not in $\bP^4$). Gauge fixing $z_5=1$
again, the four fixed points are at the solutions of
$(1+z_2^2)z_2^2 = t^2$. As we send $t\to 0$, two of these fixed points
go into the singularity, while the other two stay at finite distance,
at $z_2^2+1=0$.

\medskip

If we take the quotient of the described manifold by the involution
\eqref{eq:sigma-involution}, we would get the local structure we want
around the shrinking $S^3$, but we would have a slightly strange
behavior around the other fixed point locus, i.e. the one at
$z_2=z_5=0$. In fact, this would be a codimension-2 orientifold locus,
which is unconventional in compactifications with O3-planes. For this
reason we slightly change the base manifold, by blowing up
$\mathbb{P}^4$ along $z_2=z_5=0$. We then obtain a toric ambient space
described by the following GLSM
\begin{equation}
  \begin{array}{c|cccccc}
    & z_1 & z_2 & z_3 & z_4 & z_5 & \lambda \\
    \hline
    \bC^*_1 & 1 & 1 & 1 & 1 & 1 & 0 \\
    \bC^*_2 & 1 & 0 & 1 & 1 & 0 & 1 
  \end{array}
\end{equation}
with SR-ideal $\{z_1z_3z_4\lambda,z_2z_5\}$.
Now $z_2z_5$ is in the SR-ideal, i.e. the unwanted codimension-2 fixed point locus does not exist anymore. On the other hand, there is now a codimension-1 fixed point locus, i.e. $\lambda=0$. 
The base equation is now
\begin{equation}
  \label{eq:B-hypersurfaceblup}
  \hat{P} = \sum_{i=1,3,4} (z_5^2\lambda^2 + z_i^2) z_i^2 + (z_5^2 + z_2^2) z_2^2\lambda^4 - t^2z_5^4\lambda^4 = 0\, .
\end{equation}
It again restricts to \eqref{eq:localPnoblp} around the conifold
singularity (when $t=0$).  Notice that the fixed points
$z_1=z_3=z_4=0$ are far away from the codimension-1 fixed point locus
$\lambda=0$.  Moreover, the involution can now be given by
$\lambda \mapsto - \lambda$ (due to the scaling relations).

In order to describe $\sB=\cB/\sigma$, with
$\sigma\colon \lambda \mapsto -\lambda$, we introduce the invariant
coordinate $\Lambda=\lambda^2$. Our new base is now a complete
intersection in the ambient space described by
\begin{equation}
  \begin{array}{c|cccccc}
    & z_1 & z_2 & z_3 & z_4 & z_5 & \Lambda \\
    \hline
    \bC^*_1 & 1 & 1 & 1 & 1 & 1 & 0 \\
    \bC^*_2 & 1 & 0 & 1 & 1 & 0 & 2 
  \end{array}
\end{equation}
with SR-ideal $\{z_1z_3z_4\Lambda,z_2z_5\}$. This is a two-to-one map from the previous base, except on the fixed loci $\lambda=0$ and
$z_1=z_3=z_4=0$, where it is one-to-one. The defining equation is $\sP=0$ where $\sP$ comes
from~\eqref{eq:B-hypersurfaceblup}:
\begin{equation}
  \label{eq:B-hypersurfaceblupquot}
  \sP = \sum_{i=1,3,4} (z_5^2\Lambda + z_i^2) z_i^2 + (z_5^2 + z_2^2) z_2^2\Lambda^2 - t^2z_5^4\Lambda^2 = 0\, .
\end{equation}
As an example, consider a neighborhood of the fixed points at
$z_1=z_3=z_4=0$. We necessarily have $\Lambda\neq 0$ and $z_5\neq 0$
(otherwise at $\sP=0$ we would have $z_5=z_2=0$). We can thus gauge
fix $\Lambda=1$ and $z_5=1$, which leaves
\begin{equation}
  \label{eq:local-conifold}
  \widetilde{\sP} = \sum_{i=1,3,4}(1+z_i^2)z_i^2 + (z_2^2 + 1)z_2^2 -t^2 = 0
\end{equation}
and a $\bZ_2$ inverting the $z_1,z_3,z_4$ coordinates. So we are left
with precisely the quotient of the deformed conifold that we
had locally.

\medskip

We now uplift this configuration to F-theory. We choose the standard 
Weierstrass model: a genus-one fibration with a section,
with the fibre realised as a degree six hypersurface on
$\bP^{2,3,1}$. Notice that the base has first Chern class
$c_1(\sB)=[z_1]$.  The CY four-fold will be a complete intersection on
a an ambient toric space $\cA$ given by the GLSM
\begin{equation}
  \begin{array}{c|ccccccccc}
    & z_1 & z_2 & z_3 & z_4 & z_5 & \Lambda & x & y & z\\
    \hline
    \bC^*_1 & 1 & 1 & 1 & 1 & 1 & 0 & 0 & 0 & -1\\
    \bC^*_2 & 1 & 0 & 1 & 1 & 0 & 2 & 0 & 0 & -1\\
    \bC^*_3 & 0 & 0 & 0 & 0 & 0 & 0 & 2 & 3 & 1
  \end{array}
\end{equation}
where we took the coordinate $z$ to belong to the anticanonical bundle of $\sB$.
The degree six equation will be given by
\begin{equation}
  \sQ = x^3 + f(z_i,\Lambda)xz^4 + g(z_i,\Lambda)z^6 - y^2 = 0
\end{equation}
with $f,g$ homogeneous polynomials of the base coordinates of
degrees $(4,4)$ and $(6,6)$ respectively. 

At this point we can do a couple of sanity checks of our construction:
\begin{enumerate}
\item The neighborhood of the O3-planes on the contracting $S^3$
  should look like an elliptic fibration over $\bC^3/\bZ_2$ (with a
  fixed point) with monodromy in the fibre given by $-1\in SL(2,\bZ)$.
  In particular, it should be the case that generically $\Delta=0$
  does not intersect the location of the O3-planes. But there should
  be, at any fixed point in the base corresponding to an O3, an
  involution of the $T^2$ sending $q\to-q$ (with $q$ the flat
  coordinate in the $T^2$).
\item Relatedly, $\Delta=0$ should not intersect the conifold point in
  the base if we send $t\to 0$, so the local system is the one we
  wanted to embed originally.
\end{enumerate}

We now perform these sanity checks. We work directly in the limit
$t=0$. If we find that $\Delta\neq 0$ at the conifold point this will
imply that for finite $t$ there is no intersection with the fixed
points either (this second fact can also be checked easily
independently for $t\neq 0$, but we will not do so explicitly
here). When $t=0$, the conifold fixed point $\cP_c$, where the two O3
planes come on top of each other, is at
$(z_1,z_2,z_3,z_4,z_5)=[0\colon 0\colon 0\colon 0 \colon 1 ]$. A
generic degree $(4,4)$ section (such as $f$) restricted to $\cP_c$ has
the form
\begin{equation}
  f|_{\cP_c} = f_0 z_5^4\Lambda^2 = f_0
\end{equation}
with generically $f_0\neq 0$. So we learn that $f\neq 0$ at the fixed
points. A similar argument for $g$ gives $g|_{\cP_c}\neq 0$, and
generically also $\Delta|_{\cP_c}=4f_0^3+27g_0^2\neq 0$ for generic
$f_0$ and $g_0$. So we learn that the discriminant does not intersect
the O3-planes. Furthermore, the argument is independent of the value
of $t$, so the discriminant locus does not intersect the conifold
singularity either, and locally we just have the type IIB system of
interest. This is as expected, since locally we have a O3 involution
of a Calabi-Yau, which can be realised supersymmetrically at constant
$\tau$, so there is no need to have 7-branes to restore supersymmetry.

Let us look at the structure around a fixed point at
$z_1=z_3=z_4=0$. As discussed above, we can fix
$\Lambda=1$ and $z_5=1$. This leaves a $\bZ_2$ symmetry acting as $z\to -z$, leaving all
other non-zero coordinates invariant. Choose a root for
$\widetilde{P}=(z_2^2+1)z_2^2 - t^2$. This leaves us
with the fibre, as expected, quotiented by a $\bZ_2$ acting as
\begin{equation}
  \label{eq:sigma-fiber}
  \widehat{\sigma}\colon (x,y,z) \to (x,-y,z)\, .
\end{equation}
(We have used the $\bP^{2,3,1}$ $\bC^*$ to make the sign act on $y$,
instead of $z$.) In terms of the flat coordinate $q$ on the torus we
have $x=\wp(q)$ and $y=\wp'(q)$, using the Weierstrass
$\wp$-function. Since $\wp(q)$ is even on $q$, and thus $\wp'(q)$ odd,
we can identify the $\bZ_2$ action~\eqref{eq:sigma-fiber} precisely as
$q\to -q$, or in terms of IIB variables $(-1)^{F_L}\Omega$.

We are now going to consider the weak coupling limit, to extract a
Calabi-Yau three-fold with the wanted involution and properties. We
will also consider the simple situation in which the D7-brane tadpole
is canceled locally \cite{Sen:1996vd}. We then have $f=\phi \,h^2$
and $g=\gamma\, h^3$ where $\phi,\gamma$ are constant and $h$ is a
polynomial of degree $(2,2)$ in $z_i,\Lambda$. Its most generic form
is
\begin{equation}
  \label{eq:h-F-model}
h(z_i,\Lambda) \equiv \Lambda\, p_2(z_2,z_5) + q_2(z_1,z_3,z_4) \:, 
\end{equation}
where $p_2,q_2$ are polynomials of degree 2.
The Calabi-Yau three-fold is then given by adding the equation $ \xi^2 = h$, i.e.
\begin{equation}\label{CYeq1stEx}
\xi^2 = \Lambda\, p_2(z_2,z_5) + q_2(z_1,z_3,z_4)  \qquad {\rm AND} \qquad \sP=0
\end{equation}
in the ambient space
\begin{equation}
  \label{eq:second-IIB}
  \begin{array}{c|ccccccc}
    & z_1 & z_2 & z_3 & z_4 & z_5 & \Lambda & \xi \\
    \hline
    \bC^*_1 & 1 & 1 & 1 & 1 & 1 & 0 & 1\\
    \bC^*_2 & 1 & 0 & 1 & 1 & 0 & 2 & 1
  \end{array}
\end{equation}
with SR-ideal $\{z_2z_5,\,\, z_1z_3z_4\Lambda\xi\}$. From \eqref{CYeq1stEx} we see that we have one 
O7-plane at $\xi=0$.\footnote{
Notice that when $q_2\equiv 0$, a $\mathbb{C}^2/\mathbb{Z}_2$ 
singularity along $\xi=\Lambda=p_2$
is generated. In fact the CY is now described by
$\xi^2=\Lambda\cdot p_2$:
the orientifold divisor $\xi=0$ splits into two pieces that intersect
exactly on the $\mathbb{C}^2/\mathbb{Z}_2$ singularity. } 

The intersection form on the 5-fold ambient space is computed in the following way. Let us first take the basis $D_1=D_{z_1}$ and $D_2=D_{z_2}$. We moreover observe that we have one point at $(z_1, z_2, z_3, z_4, z_5, \Lambda, \xi)=(0,0,0,0,1,0,1)$, i.e. $H_1^4(2H_1-2H_2)=1$. The SR-ideal tells us that $H_2^2=0$ and $H_1^4(H_1-H_2)=0$. Hence we obtain that the only two non-vanishing intersection numbers are $H_1^5=H_1^4H_2=\frac12$. Hence on the CY three-fold, that is defined by intersecting the two divisors in the classes $[P]=4H_1$ and $[\xi^2]=2H_1$, we have the intersection form
\begin{equation}
  I_3 = 4 D_1^3 + 4 D_1^2D_2 \:.
\end{equation}
We can also compute the second Chern class of the three-fold by adjunction. We obtain
\begin{equation}
c_2(X_3) = 6D_1^2 + 4D_1D_2 \:.
\end{equation}

For our purposes, the important equation is $\sP=0$ that gives the
conifold singularity and the physics we are interested in. One can
easily see that in the double cover description we indeed have a
conifold singularity (as opposed to its quotient). To see this, zoom
on the neighborhood of $z_1=z_3=z_4=0$. Looking to the
expression~\eqref{eq:h-F-model}, we see that on the $z_1=z_3=z_4=0$
locus $\Lambda=0$ implies $\xi=0$, so given that $z_1z_3z_4\Lambda\xi$
belongs to the SR-ideal, we conclude that in this neighborhood, for
generic $p_2(z_2,z_5)$, $\Lambda\neq 0$. We can thus gauge fix
$\Lambda=1$, which leaves a $\bZ_2$ subgroup unfixed. This subgroup is
precisely the one that exchanges the two roots of $\xi^2=h$, so we can
gauge fix it by choosing arbitrarily one of the roots in the whole
neighborhood. As above, due to the $\sP=0$ equation and the fact that
$z_2z_5$ belongs to the SR-ideal, we have that $z_5\neq 0$, so we can
use it to gauge-fix the remaining $\bC^*$ symmetry by setting
$z_5=1$. We thus end up with $\widetilde{P}(z_1,z_2,z_3,z_4)=0$, as
in~\eqref{eq:local-conifold}, i.e. a deformed conifold singularity
with no quotient acting on it.

We can also compute the Euler characteristic of the O7-plane divisor,
that allows us to compute the D3-charge of the O7-plane and four
D7-branes (plus their images) on top of it. In fact we have
$[O7]=H_1$. The Euler characteristic of a four-cycle $D$ is
\begin{equation}
 \chi(D)=\int_D c_2(D) = \int_{X_3} D(D^2+c_2(X)) \:,
\end{equation}
where in the last step we used the adjunction formula (to obtain $c_2(D)=c_1(D)^2+ c_2(X)$) and the fact that $D$ is a divisor of a CY (and then $c_1(D)=-D$).
In our case $\chi([O7])=H_1^3+H_1c_2(X) = 44$. Hence the geometric induced D3-charge of the system made up of the O7-plane plus four D7-branes (plus their images) on top of it is 
\begin{equation}
Q_{D3}^{(4D7+O7)} = - \frac{\chi([O7])}{2} = -22\:.
\end{equation} 
Any half-integral flux (that could be induced by the Freed-Witten anomaly cancellation condition) gives an integral contribution in this configuration (since there are eight D7-branes).

\medskip

One further thing to check is whether there are constraints on the
NSNS three-form flux coming from the Freed-Witten anomaly
\cite{Witten:1998xy,Freed:1999vc}. These will be absent if
$H^3(O7, \bZ)=0$. This follows from the Lefschetz hyperplane theorem
as follows. We start by desingularising the ambient toric
space~\eqref{eq:second-IIB}. The singularity is at
$z_1=z_3=z_4=\xi=0$. It can be easily seen that this locus does not
intersect the Calabi-Yau hypersurface, for generic $p_2,q_2$. So if we
blow up along the singular point we do not alter the Calabi-Yau
itself (or any of its divisors). A possible desingularised ambient
space is
\begin{equation}
  \label{eq:desingularized-A}
  \begin{array}{c|cccccccc}
    & z_1 & z_2 & z_3 & z_4 & z_5 & \Lambda & \xi & s \\
    \hline
    \bC^*_1 & 1 & 1 & 1 & 1 & 1 & 0 & 1 & 0\\
    \bC^*_2 & 1 & 0 & 1 & 1 & 0 & 2 & 1 & 0\\
    \bC^*_3 & 0 & 0 & 0 & 0 & 0 & 1 & 0 & 1
  \end{array}
\end{equation}
with SR-ideal $\{z_2z_5, s\Lambda, z_1z_3z_4\xi\}$. The locus of
interest is given by
\begin{equation}
  \{\xi = 0\} \quad \cap\quad \{s\xi^2 = \hat{h}\} \quad \cap \quad \{\hat{P}=0\}
\end{equation}
with
\begin{equation}
  \hat{P} = \sum_{i=1,3,4} (z_5^2\Lambda + sz_i^2) sz_i^2 + (z_5^2 + z_2^2) z_2^2\Lambda^2 - t^2z_5^4\Lambda^2
\end{equation}
and
\begin{equation}
  \hat{h} = \Lambda p_2(z_2,z_5) + s q_2(z_1,z_3,z_4)
\end{equation}
the proper transforms of the original divisors. We start by imposing
$\xi=0$. This gives rise to a toric space $\cA_\xi$ of one dimension
lower, which can easily be seen to be smooth. Similarly, $\hat{h}=0$
gives rise to a smooth hypersurface $Y$ in $\cA_\xi$, and it can be
seen that the O7 locus $\hat{P}=0\subset Y$ is also smooth. So by
straightforward repeated application of the Lefschetz hyperplane
theorem we learn that $H_1(O7,\bZ)=H_1(\hat{\cA},\bZ)$, with
$\hat{\cA}$ the ambient toric space~\eqref{eq:desingularized-A}. But
it is easy to see that $\pi_1(\hat{\cA})=0$ from standard
considerations in toric geometry (see for instance theorem 12.1.10
in~\cite{cox2011toric}), so by the Hurewicz isomorphism and Poincaré
duality on the O7 worldvolume we learn that $H^3(O7,\bZ)=0$.

\subsection{Goldstino retrofitting}\label{Sec:secondEXAMPLE}

The model in the previous section was designed in order to display the
structure of interest. While this is interesting, it is also
interesting to see if existing, phenomenologically interesting type IIB
models with O3-planes admit the addition of a nilpotent Goldstino
sector, ``retrofitting'' them with a possible de Sitter uplift
mechanism at little cost.

To show that this is indeed the case, we consider the model in
\cite{Diaconescu:2007ah,Cicoli:2012vw}. It is constructed starting
from a hypersurface in the toric ambient space
\begin{equation}
\begin{array}{c|cccccccc|c}
 & W_1 & W_2 & W_3 & W_4 & W_5 & Z & X & Y & D_\textmd{H} \tabularnewline \hline 
    \bC_1^* & 0  &  0  &  0  &  0  &  0  &  1  &  2  &  3  & 6\tabularnewline 
    \bC_2^* & 1  &  1  &  1  &  0  &  0  &  0  &  6  &  9  & 18\tabularnewline
    \bC_3^* & 0  &  1  &  0  &  1  &  0  &  0  &  4  &  6  & 12\tabularnewline
    \bC_4^* & 0  &  0  &  1  &  0  &  1  &  0  &  4  &  6  & 12\tabularnewline
\end{array}\label{eq:model3dP8:weightm}\,,
\end{equation}
with SR-ideal
\begin{equation}
\label{eq:model3dP8:sr-ideal}
{\rm SR}=\{W_1\, W_2\,W_3,\, W_2\, W_4,\, W_3 \, W_5,\, W_4\, W_5, \, W_1\,W_2\,X\,Y, \, W_1\,W_3\,X\,Y, \,  W_4\,Z, \,W_5\,Z, \, X\, Y\, Z\}\,.
\end{equation}
The last column indicates the degree of the polynomial defining the CY
three-fold. This polynomial takes the form of a Weierstrass model
\begin{equation}\label{CYeq2ndMd}
 Y^2 = X^3 + f(W_i) \,X\,Z^4 + g(W_i)\,Z^6 \:,
\end{equation}
where $f$ and $g$ are respectively polynomials of degree $(0,12,8,8)$ and $(0,18,12,12)$ in the coordinates $W_1,\ldots,W_5$. 

This CY $X$ has Hodge numbers $h^{1,1}=4$ and $h^{1,2}=214$.
The intersection form takes the simple expression
\begin{equation}
  I_3 = 9D_1^3 + D_2^3 + D_3^3 + 9D_4^3 
\end{equation}
in the following basis of $H^{1,1}(X)$:\footnote{This is not an integral basis: for example $D_{W_1}=\frac16(D_1-3D_2-3D_3-D_4)$.}
\begin{equation}\label{dP8model-basis}
 D_1 = 3D_{W_3} + 3D_{W_4} + D_{Z} \qquad D_2 = D_{W_4} \qquad D_3 = D_{W_5} \qquad D_4 =  D_{Z} \:.
\end{equation}
Three of the basis elements are del Pezzo surfaces. In particular $\{Z=0\}$ is a $dP_0$, while $\{W_4=0\}$ and $\{W_5=0\}$ are $dP_8$'s.  
The second Chern class of the Calabi-Yau is
\begin{equation}\label{c2CY2ndExample}
c_2(X_3) = \frac{1}{3}\left(34 D_1^2 + 30 D_2^2 + 30 D_3^2 - 2 D_4^2 \right)\: .
\end{equation}
We consider the involution \cite{Diaconescu:2007ah,Cicoli:2012vw} 
\begin{equation}
 W_2 \leftrightarrow W_3 \qquad \mbox{and} \qquad W_4 \leftrightarrow W_5 
\end{equation}
exchanging the two $dP_8$'s. 
The CY three-fold equation must be restricted to be invariant under this involution. $X,Y,Z$ are invariant under such involution. The rest of invariant monomials are $W_1$, $u\equiv W_2W_3$, $v\equiv W_4W_5$ and $w\equiv W_3W_4+W_2W_5$. The equation becomes invariant if $f$ and $g$ depend on $W_i$ only as functions of $W_1,u,v,w$.

Let us consider the fixed point locus. It is made up of a codimension-1 locus at $W_3W_4-W_2W_5=0$ and four isolated fixed points: one at the intersection $W_3W_4+W_2W_5=W_1=Z=0$ and three at the intersection $W_3W_4+W_2W_5=W_1=Y=0$~\cite{Diaconescu:2007ah,Cicoli:2012vw}. So by implementing this orientifold involution, one obtains one O7-plane in the class $[D_{O7}]=[D_{W_3}]+[D_{W_4}]$ and four O3-planes.

We focus on the neighborhood of the O3-planes at $Y=W_1=w=0$ (we have
used the above definition $w\equiv W_3W_4+W_2W_5$). If we plug these
relations inside the defining equation \eqref{CYeq2ndMd}, we get a
cubic in X
\begin{equation} \label{cubicO3pl}
X^3 + \alpha X u^6 v^2 Z^4 + \beta u^9 v^3 Z^6 = 0
\end{equation}
where, as said above, $f$ and $g$ are functions of the invariant
monomials, and $\alpha,\beta$ are tunable complex structure moduli.
First of all, because of SR-ideal, we know that $u$, $v$ and $Z$ are
non-vanishing.\footnote{We have $W_1=0$. $u=0$ would mean either
  $W_2=0$ or $W_3=0$. In the first case, $w=0$ would mean either
  $W_3=0$ or $W_4=0$; but both $W_1W_2W_3$ and $W_2W_4$ are in the
  SR-ideal. The same considerations are valid for $W_3=0$. Hence,
  $u=0$ cannot be realised on this locus. The same conclusions are
  valid for excluding the intersection with $v=0$. $Z\neq0$ is even
  easier: since $Y=0$, $Z$ cannot be zero as well, otherwise the
  equation would give $X=0$ too and $XYZ$ is in the SR-ideal.}  We can
thus fix $W_4=W_5=Z=1$ and $W_2=i$ via the projective rescalings, in
which case $W_3W_4+W_2W_5=0$ becomes simply $W_3=-i$. In terms of the
invariant coordinates we have $u=v=1$. With this gauge choice we have
that \eqref{cubicO3pl} becomes
\begin{equation}
  X^3 + \alpha X + \beta = 0\, .
\end{equation}
Hence the zeros of \eqref{cubicO3pl} are at the zeros of the cubic
equation. We are interested to the case when two of these zeros come
together. This happens when the discriminant of the cubic is zero,
i.e. when
\begin{equation}
 \Delta\equiv 4 \alpha^3 + 27 \beta^2 = 0 \:,
\end{equation}
that is a relation among the complex structure parameters. We can parametrise this situation by taking $\alpha=-3a^2$ and $\beta=2a^3-\delta$. When $\delta=0$ two of the roots come together. We can also rewrite the cubic equation as
\begin{equation} \label{cubicO3pl2}
(X-a)^2(X+2a) - \delta  = 0 \:.
\end{equation}
Now it is manifest that when $\delta=0$ we have a double root at $X=a$.

Let us study the local form of \eqref{CYeq2ndMd} around
$Y=W_1=w=X-a=0$. As above, we use the $\bC^*$ symmetries to fix
$W_4=W_5=Z=1$. In addition, we gauge fix $u=1$. Notice that this
leaves an unfixed $\bZ_2$ subgroup, generated by
$(\lambda_1,\lambda_2,\lambda_3,\lambda_4)=(1,-1,1,1)$. We choose to
fix this subgroup by requiring that at the fixed point $W_2=i$, as
above, or in terms of coordinates on a neighborhood of the point, that
$W_2=i+\omega-\frac{i}{2}\omega^2$, with $|\omega|\ll 1$.\footnote{For
  each $W_2$ there are two values for $\omega$, but only one of these
  values satisfies $|\omega|\ll 1$ for $|W_2-i|\ll 1$, so we can
  consistently choose this value to define a one-to-one map between
  $W_2$ and $\omega$ in a neighborhood of the conifold.} To quadratic
order in $\omega$ this gauge fixing implies
$W_3=W_2^{-1}=-i+\omega+\frac{i}{2}\omega^2$.  Expanding in terms of
these new coordinates around $Y=W_1=w=X-a=0$ we have
\begin{equation}
  -Y^2 + (X-a)^2 (3a+\ldots)  + W_1^2 (c_{W_1}+\ldots) + \omega^2 (c_\omega+\ldots) = \delta
\end{equation}
where $\ldots$ are terms that vanish on the analysed locus and
$c_{W_1}$ and $c_w$ are generically non zero constants (in the chosen
patch). We have used the freedom in redefining the complex coordinates
in order to erase possible $W_1\omega$ mixed terms. We immediately
see that we obtain a conifold singularity when $\delta\rightarrow 0$.

How does the permutation involution act on this local conifold? The
coordinates $X,Y,Z,u,v$ are all invariant, as is the gauge fixing
$Z=W_4=W_5=u=1$. On the other hand, the image of
$(W_2,W_3)=(i+\omega-\frac{i}{2}\omega^2,
-i+\omega+\frac{i}{2}\omega^2)$
is
$(W_2,W_3)=(-i+\omega+\frac{i}{2}\omega^2,i+\omega-\frac{i}{2}\omega^2)$,
which is not in the form given by the $\bZ_2$ gauge fixing above. We
can go back to the desired gauge frame by acting with $\lambda_2=-1$,
which acts on our local conifold coordinates as
\begin{equation}
  (X-a,Y,W_1,\omega) \mapsto (X-a, -Y, -W_1, -\omega)
\end{equation}
and perfectly reproduces the geometric action required for the
retrofitting of a nilpotent Goldstino sector.

\medskip

Let us finish with some considerations on the D3-charge. Remember
first that we have four O3-planes. The D3-charge of the system of the
O7-plane and four D7-branes (plus their images) on top of it, is given
by $-\chi([O7])/2$, where $[O7]$ is the homology class of the O7-plane
locus. In our case $[O7]=D_{W_3}+D_{W_4}=\frac13(D_1-D_4)$. By using
\eqref{c2CY2ndExample}, the Euler characteristic can be computed, as
$\chi(D)=D^3+c_2(X_3)D$ in a CY three-fold. We obtain
$\chi([O7])=36$. Hence the localised objects in the compactification
have integral D3-charge. As discussed in \S\ref{sec:decay}, choose for
instance to put a stuck D3 at one of the O3$^-$ points on the
contracting $S^3$, and a stuck \ad3 on the other O3$^-$ on this same
$S^3$. This pair of stuck branes does not contribute to the D3
tadpole. Finally, recall that we introduce fractional branes in the
orientifolded conifold cascade in order to create the warped throat by
confinement. The number of branes to introduce is arbitrarily tunable,
and completely determines the amount of D3 charge induced by the
fluxes in the confined description, i.e. threading the warped
throat.

\section{Conclusions}\label{Sec:conclus}

We have presented the first explicit CY compactifications with anti-D3-branes at the tip of a long throat for which the single propagating degree of freedom is the goldstino and therefore can be represented by a nilpotent superfield.
 
Anti-D3-branes are an important tool for type IIB phenomenology. An anti-D3-brane 
at the tip of a warped throat, generated by three-form fluxes \cite{Giddings:2001yu}, produces an uplifting 
term to the scalar potential, that allows to obtain de Sitter minima \cite{Kachru:2003aw}. 
By a mild tuning of the three-form fluxes, one can get a fine tuning of the 
cosmological constant, that is model independent 
if the throat is localised far from the visible spectrum. The presence of the 
anti-D3-brane can be described in a supersymmetric effective field theory 
(even if non-linearly) by the introduction of constrained superfields.
The  simplest situation is when the anti-D3-brane is on top of an O3-plane 
at the tip of the throat: one needs just to add one nilpotent superfield that 
captures the goldstino degrees of freedom. This has been studied in \cite{Kallosh:2015nia}.

In this paper we have discussed how to realise this setup in a
globally consistent Calabi-Yau compactification.  The necessary
ingredients are a warped throat, realised by considering a KS deformed
conifold throat embedded in a compact CY like in \cite{Giddings:2001yu}, and an
orientifold involution that produces a couple of O3-planes at the tip
of the throat.

We first analysed the local neighbourhood of the O3/$\ad3$ system.  We
started from considering the conifold singularity. It is well know
that putting three-form fluxes on the deformed conifold produces a
warped throat with a three-sphere at the tip.  This three-sphere
collapses when the deformation goes to zero and the conifold
singularity is generated. We have first studied the situation for the
singular conifold and then transported our result to the deformed
one. We have considered the simplest involution that keeps the
singularity fixed. This involution has no fixed points in the resolved
phase (although this statement is somewhat subtle due to the fact that
the geometric resolution mode is projected out, as we have explained),
but has still two fixed points on the deformed phase, that are placed
on the north and south poles of the three-sphere at the tip of the
throat. These two fixed points collapse on top of each other when one
takes the singular limit. Hence, by using this orientifold involution
in the deformed phase, one generates two O3-planes at the tip of the
throat.  We also mapped the system to the T-dual type IIA
configuration, that is well known also in the orientifolded case. This
allowed us to double check some of our conclusions and solve some
apparent puzzles.

For the unorientifolded KS throat it is well known that the deformed
phase is realised dynamically in the field theory living on a stack of
D3-branes probing the conifold singularity: the classical moduli space
is deformed quantum mechanically due to the dynamically generated
F-terms. The same process takes place in the orientifolded case, and
by a careful analysis of the quantum dynamics of the $SO\times \Sp$
theory at the singularity we have determined which type of O3-plane is
generated. We have found agreement with the prediction from the type
IIA dual configuration: the two O3-planes are of the same type, either
both $O3^-$ for $\Sp$ confinement (with one or both of type
$\widetilde{O3}^-$), or both $O3^+$ for $SO$ confinement.

We have used the local results outlined above to embed the system in a
compact CY.  We have found two examples. In both cases, we have
constructed a CY three-fold with the following properties: 1) It has a
definite complex structure deformation that allows to take the
explicit conifold limit, i.e. we have identified a parameter in the CY
defining equation that generates a conifold singularity when set to
zero.  2) It has an involution that, in the local patch around the
conifold singularity (or the tip of the deformed one), acts in the
same way we found in the local analysis and that gives two O3-planes
on top of the deformed conifold three-sphere. We have followed two
procedures to find our compact examples. In the first case we
constructed the CY, by first embedding the orientifolded conifold in a
non-Calabi-Yau compact threefold. Then, by constructing a F-theory
model over this base and taking its Sen weak coupling limit, we have
constructed a CY three-fold with the wanted features. In the second
case, we started with a previously studied phenomenologically
interesting CY with an involution that generates more than one
O3-plane and then checked that there is a deformation of the defining
equation that brings two O3-planes on top of each other. We showed
that this deformation generates a conifold singularity on the point
where the two O3-planes coincide.  It would be interesting to
systematise both methods (direct construction and search) to obtain a
list of suitable CYs.

\medskip

\begin{figure}
\centering
\includegraphics[width=10.0cm]{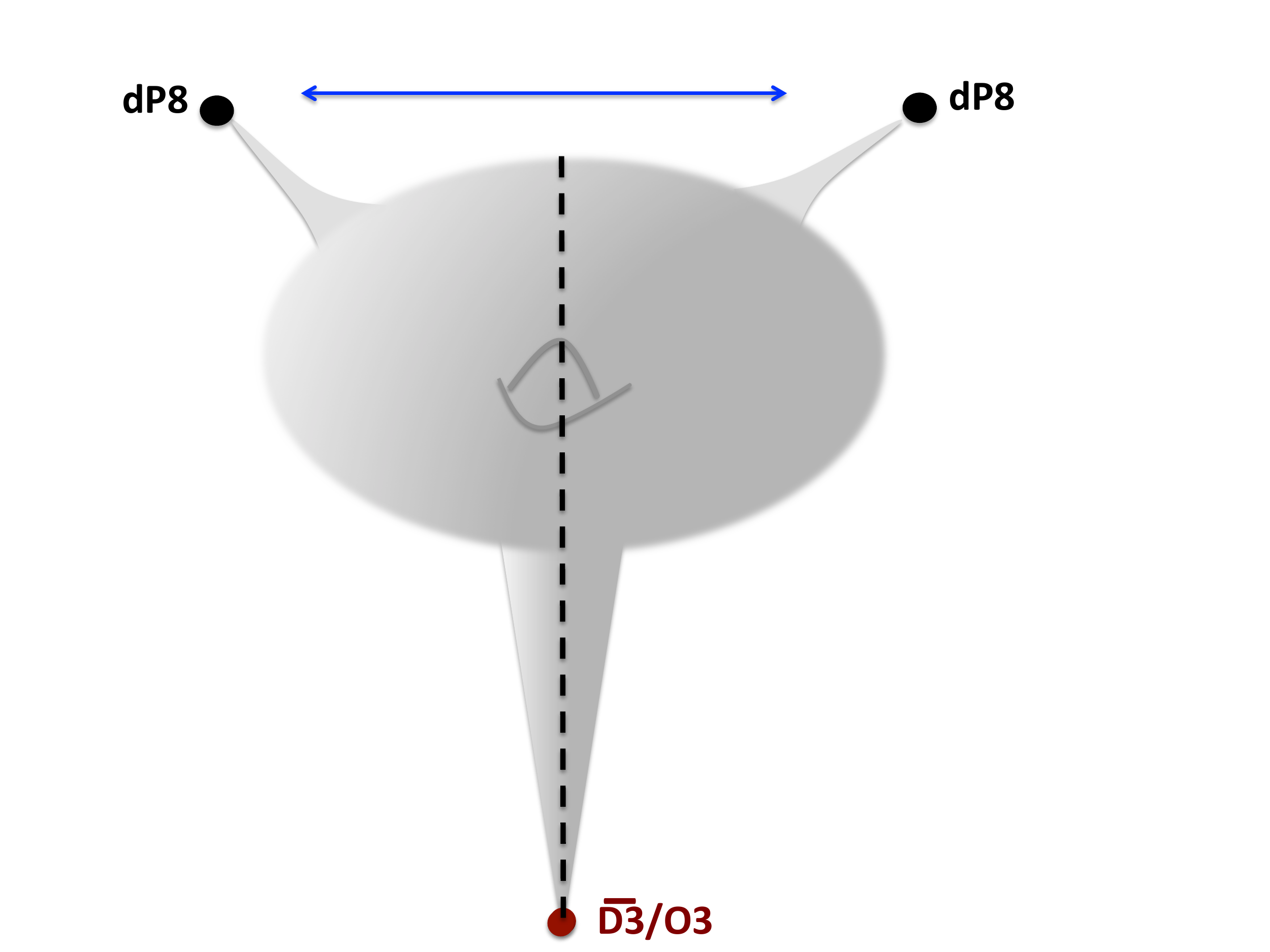}
\caption{CY manifold of \S\ref{Sec:secondEXAMPLE}, when two dP$_8$ divisors are shrunk to zero size to generate two singularities (exchanged by orientifold involution). D3-branes on these singularities produce non-abelian gauge groups and chiral spectrum.}
\label{Fig2}
\end{figure}

In summary, we have achieved the concrete construction of simple
models satisfying all requirements for a proper global embedding of
the $\ad3$ at the tip of a throat with the nilpotent goldstino in the
spectrum. One can extend our results to present explicit calculations
for moduli stabilisation for both KKLT \cite{Kachru:2003aw} and Large Volume (LVS)
\cite{Balasubramanian:2005zx,Conlon:2005ki} scenarios.  The CY
manifold in \S\ref{Sec:secondEXAMPLE} and the studied involution was
used in \cite{Cicoli:2012vw} to realise a type IIB global model with
chiral spectrum coming from D3-branes at dP$_8$ singularities (these
are realised by shrinking the four-cycles $D_{W_4}$ and $D_{W_5}$) and
with all geometric moduli stabilised.  In that paper the dS uplift was
meant to be induced by a T-brane \cite{Cicoli:2015ylx}, but here we
have shown that also the anti-D3-brane uplift may be realised. See
figure \ref{Fig2} for a picture of the setup.  We will study this
example in detail in a future work.

\bigskip

\acknowledgments

We thank Michele Cicoli and \'Angel Uranga for useful conversations.

\bigskip

\bibliographystyle{JHEP}
\bibliography{refs}

\end{document}